\pdfoutput=1

\documentclass{article}

\usepackage{microtype}
\usepackage{graphicx}
\usepackage{multirow}
\usepackage{amsmath}

\usepackage{xcolor} 

\usepackage{hyperref}
\usepackage{xurl}
\usepackage{booktabs,pifont,tabularx,array}
\usepackage{placeins}

\newcommand{\tm}{\texttrademark}
\newcommand{\sysname}{KOSMOS}
\newcommand{\sysposs}{KOSMOS'}
  \usepackage[accepted]{mlsys2025}
  \makeatletter\renewcommand{\Notice@String}{Preprint.}\makeatother
\usepackage{rotating}

\newcommand{\cmark}{\ding{51}}
\newcommand{\xmark}{\ding{55}}

\mlsystitlerunning{\sysname{}}

\begin{document}

\twocolumn[
\mlsystitle{KOSMOS: Self-scheduling Persistent Kernels for Communication-Compute Overlap}

\begin{mlsysauthorlist}
\mlsysauthor{Alexander Magro}{AMD}
\mlsysauthor{Aristotle Martin}{AMD}
\mlsysauthor{Ye Wang}{AMD}
\mlsysauthor{Wen Chen}{AMD}
\end{mlsysauthorlist}

\mlsysaffiliation{AMD}{Advanced Micro Devices Inc., Austin, TX, USA}

\mlsyscorrespondingauthor{Alexander Magro}{alex.magro@amd.com}

\mlsyskeywords{Mixture-of-Experts, Tensor Parallelism, Sequence Parallelism, Megakernels, Persistent Kernels, Distributed LLM Training, AMD GPUs, Communication-Computation Overlap}

\vskip 0.3in

\begin{abstract}
Tensor and sequence parallelism (TP+SP) and expert parallelism (EP) put collectives on the critical path of transformer layers. State-of-the-art fused kernels mitigate this with fixed roles or schedules, but this rigidity leaves large idle windows. Predicting a communication-computation split does not remove them either; a predetermined split cannot follow data readiness. We present \sysname{}, a library for self-scheduling persistent kernels whose workgroups use queues ordered by consumer demand to switch between communication and computation roles as data become ready, with no scheduler, per-call cost model, or tuned split. \sysname{} fuses every TP+SP collective of a transformer layer with its GEMM, and runs an entire expert layer, from router to fold, in one launch per direction. Evaluated on AMD Instinct\tm{} MI350X and AMD Instinct\tm{} MI355X GPUs, \sysname{} speeds up TP+SP end-to-end training in Megatron-LM by 1.10--1.28$\times$ in BF16 and 1.10--1.25$\times$ in MXFP8. \sysname{} accelerates end-to-end training of DeepSeek-V3 and Qwen3-235B proxies by 1.29--1.83$\times$ over DeepEP.
\end{abstract}
]

\printAffiliationsAndNotice{}

\section{Introduction}

Training frontier large language models (LLMs) requires parallelism strategies that trade memory for communication. Memory reduction strategies such as tensor and sequence parallelism (TP+SP) and expert parallelism (EP) in Mixture-of-Experts (MoE) put collectives on the critical path. TP+SP wraps every transformer block in all-gather (AG) and reduce-scatter (RS) collectives, whereas EP in MoE models adds all-to-alls (A2A) around each expert layer. To keep matrix cores and links saturated, overlapping communication and computation is critical to training throughput.

Fusing communication into general matrix multiplication (GEMM) avoids hardware-scheduler contention and moves data at tile granularity, overlapping with the GEMM's tile-based computation.
Prior fused kernels dedicate a fixed number of compute units (CUs) to communication~\cite{MLSYS2025_c6ee784c,awad2025irisfirstclassmultigpuprogramming,MLSYS2026_ff997469} or assign each workgroup (WG) its tasks with a scheduler~\cite{NEURIPS2025_918d938b,318585}. 
However, communication demand is not fixed: In dense collectives it shifts as the kernel progresses, and in MoE it shifts with each batch's routing.
Predicting a split of communication and computation WGs does not solve this issue. A planned split from a cost model before each call is slower than a fixed split in our measurements (Table~\ref{tab:policy}). Even adapting splits at runtime, as mKernel does with a kernel controller, leaves their kernel 1.18$\times$ slower than the best fixed split~\cite{mao2026mkernelfastmultigpumultinode}.

We present KOSMOS (Kernel-Overlapped Self-scheduling Megakernels On Symmetric-memory), a library of persistent kernels in which WGs dynamically change roles as the kernel runs. A WG is either a mover, which mainly communicates, or a builder, which mainly computes GEMM tiles. A WG selects its role from data readiness, with no scheduler or per-call cost model. Scheduling is implicit: Data is produced in the order consumers claim it. Builders start on local data while movers gather; reductions are produced, ordered by groups of tiled rows (band-major), so every rank's movers start after one step, and dispatch delivers experts in the order each GEMM consumes them. 

\begin{sidewaysfigure*}
\centering
\includegraphics[width=\textheight]{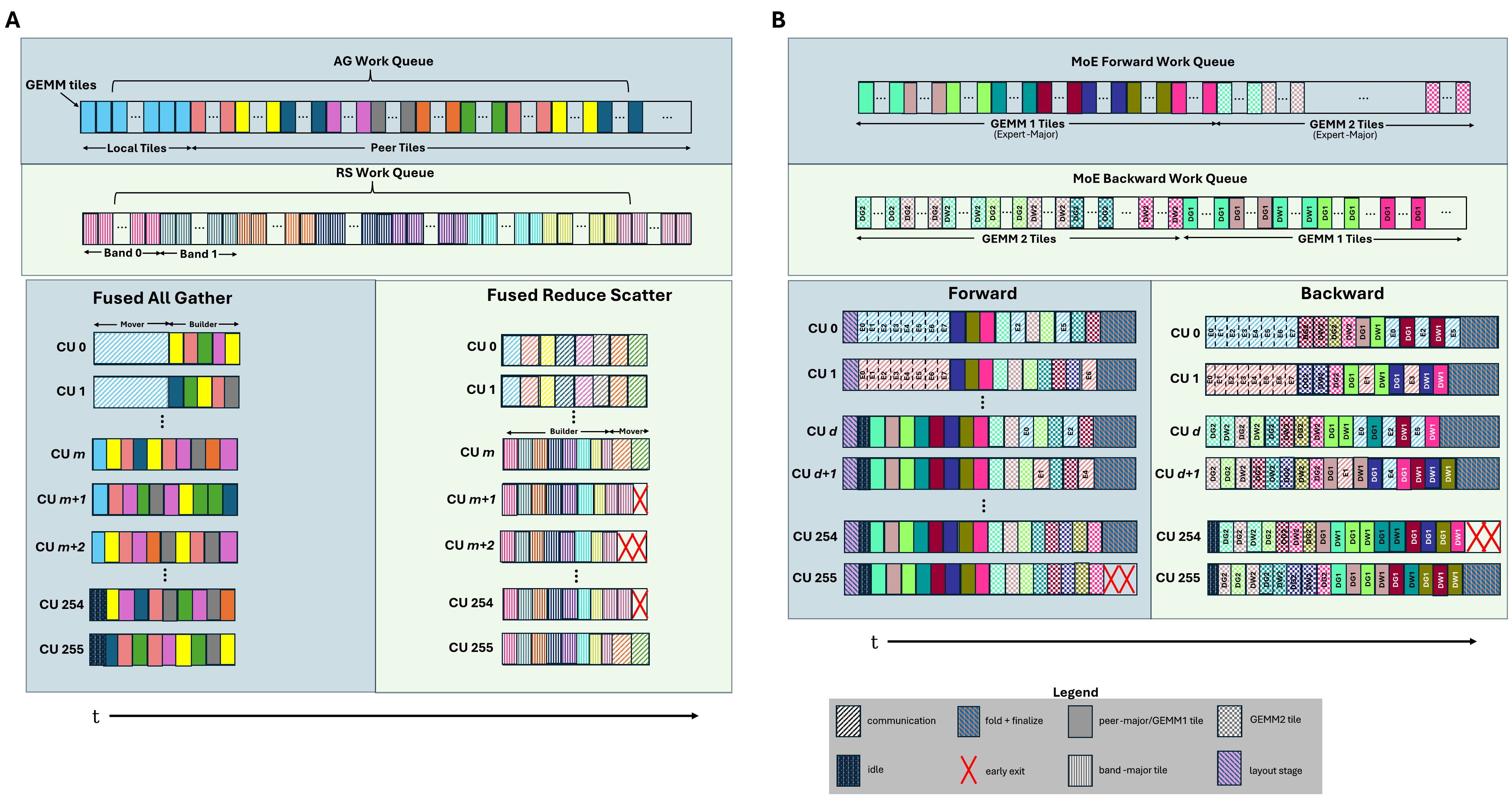}%
\caption{CU timelines and work queues on eight GPUs for (A) fused AG and RS and (B) the MoE forward and backward. Each CU runs one persistent WG; queues hold GEMM output tiles, and hatched segments are transfers (color = rank or expert). (A) The AG queue holds local tiles first, then peer tiles interleaved row by row; the RS queue is band-major (pattern = band). AG movers (CUs 0--1 of $m$) become builders once their transfers finish; in RS, builders become movers when the GEMM queue empties. (B) The forward queue is expert-major, GEMM1 before GEMM2; the backward queue holds GEMM2 then GEMM1 gradient tiles, interleaving $dG$ and $dW$. All CUs first route and plan (purple); dispatchers push, then build; combine movers (CUs $d$, $d+1$) build until a combine unit is ready, then interleave pushes with GEMM2 tiles. Backward: $dY$ is dispatched first and the $dX$ combine overlaps the GEMM1 gradients.}\label{fig:cu_timeline}
\end{sidewaysfigure*}

The switching policy is set by each collective-GEMM pair (Figure~\ref{fig:cu_timeline}). In AG+GEMM and MoE dispatch, movers become builders once their transfers are done; in GEMM+RS (Figure~\ref{fig:cu_timeline}(A)), builders join the reduction when the GEMM work queue is exhausted; in MoE combine (Figure~\ref{fig:cu_timeline}(B)) WGs switch both ways, pushing whenever data is ready and building otherwise. The policy is designed around the shape of the demand: Two-way switching is slower for TP+SP and A2A dispatch collectives whose demand is monotonic; one-way switching leaves combine gaps larger than needed.

\sysname{} provides broad coverage, including, to our knowledge, the first readiness-driven TP+SP fused kernels and MoE megakernels for training. Our contributions are:
\begin{enumerate}
    \item \textbf{Self-scheduling persistent kernels}, with producer and consumer queues co-designed per collective around data readiness. WGs may then switch roles from local tile-level checks. Switching lets a shape-only heuristic replace per-shape tuning: within 1\% of, and up to 15\% faster than, the best single fixed split. A fixed split in turn beats a split predicted per call.
    \item \textbf{A characterization of where idle time falls} in fused communication kernels and a policy rule that reduces it: one-way switching for monotonic demand, two-way switching for interleaved demand.
    \item \textbf{Complete kernel coverage for training:} persistent fused kernels for every TP+SP collective, forward and backward, in BF16 and MXFP8, and the first role-switching MoE layer with a single launch per direction, from router to fold, including gradients, evaluated end-to-end in Megatron-LM on two GPU models.
    \item \textbf{Communications optimized for bytes/CU/s:} collectives for fully connected meshes that saturate the bandwidth while maintaining fine grained access, and measurements showing that on power-capped GPUs the CUs lent to communication cost the GEMM little.
\end{enumerate}
We demonstrate \sysname{} on MI350-series GPUs and expect many of its findings to generalize to other systems: On AMD Instinct\tm{} MI350X and AMD Instinct\tm{} MI355X GPUs, \sysname{} speeds up end-to-end Llama~3 training by up to 1.28$\times$ over Transformer Engine (TE), and its MoE layers speed up training by up to 1.83$\times$ over DeepEP.

\section{Background and Motivation}
\subsection{Parallel Strategies}
\subsubsection{Tensor and Sequence Parallelism}
Under TP+SP~\cite{shoeybi2020megatronlmtrainingmultibillionparameter,MLSYS2023_80083951}, each transformer block contains a column-parallel and a row-parallel GEMM pair (Figure~\ref{fig:tpsp_flowchart}). Activations outside the pair are sequence-sharded. An AG assembles the sequence before a column-parallel GEMM, and an RS after the row-parallel GEMM reduces sequence shards. The backward pass adds three collectives in BF16: AGs of the output gradients and the saved input, and an RS of the input gradient. MXFP8 gathers output gradients twice, quantized along both dimensions. Each GEMM pair runs five or six collectives per training step, each on the critical path unless overlapped.

\subsubsection{Fully Sharded Data Parallelism}
Fully sharded data parallelism (FSDP) shards only the model states, allowing collectives to gather model states for the next layer to run off of the critical path. Intra-node FSDP~\cite{zhao2023pytorchfsdpexperiencesscaling,9355301} is often been preferred to TP+SP on AMD Instinct nodes~\cite{rocm_megatron_training_doc,11520488}. However, for large-scale distributed training, FSDP is often paired with TP+SP, which then shards activation memory across sequence, head, and channel dimensions, which dominates at scale.

\subsubsection{Expert Parallelism}
In an MoE layer with EP, the experts are distributed throughout the world size. A router GEMM matches $k$ experts to each token, then an A2A dispatch sends activations to the selected experts' GEMMs. An A2A combine then returns the expert outputs to the token's original rank, with results folded by token. Because routing is decided per pass, transfer volume and destinations change every step, with skew towards GPUs hosting hot experts. The more skewed the data, the more completion times diverge.

\subsection{MXFP8 Training}
MXFP8~\cite{rouhani2023microscalingdataformatsdeep} stores each element in FP8 and shares one power-of-two (E8M0) scale across each block of 32 consecutive elements, so a tensor moves roughly half the bytes of BF16. A block is quantized in the contracting direction. In training, as each matrix and its transpose are used, we quantize both row-wise and column-wise. Requantizing one copy from the other compounds rounding error, so both must be quantized from the higher-precision source. MXFP8 has been shown to be as accurate in training as BF16~\cite{mishra2025recipespretrainingllmsmxfp8}, with twice the theoretical GEMM throughput on MI350-series GPUs.

\subsection{Motivation and Design Requirements}\label{sec:motivation}

For TP+SP, the common overlap strategy pipelines chunks of data through transport and compute on two streams. This is generally inefficient, because the two kernels compete for the same CUs without coordinating. In an eight-GPU node in the MI350-series, each GPU has a theoretical bidirectional peak of 1008 GB/s of bandwidth~\cite{amd_mi350x}, split between seven 144 GB/s XGMI links. A single peer is reachable at 1/7th of the aggregate. Full bandwidth usage requires sending to and receiving from all peers simultaneously. Host initiated direct memory access engines cannot provide this fan-out at fine granularity: They lag for small transfers because of their launch and synchronization cost, and they cannot perform reductions needed by TP+SP or EP. The tile-granular overlap therefore requires in-kernel communication. This leads to our first requirement: \textbf{R1: One kernel, one stream:} Communication and computation must share the GPU within one kernel.

\begin{figure*}[p]
    \centering
    \includegraphics[width=\linewidth,height=0.40\textheight,keepaspectratio]{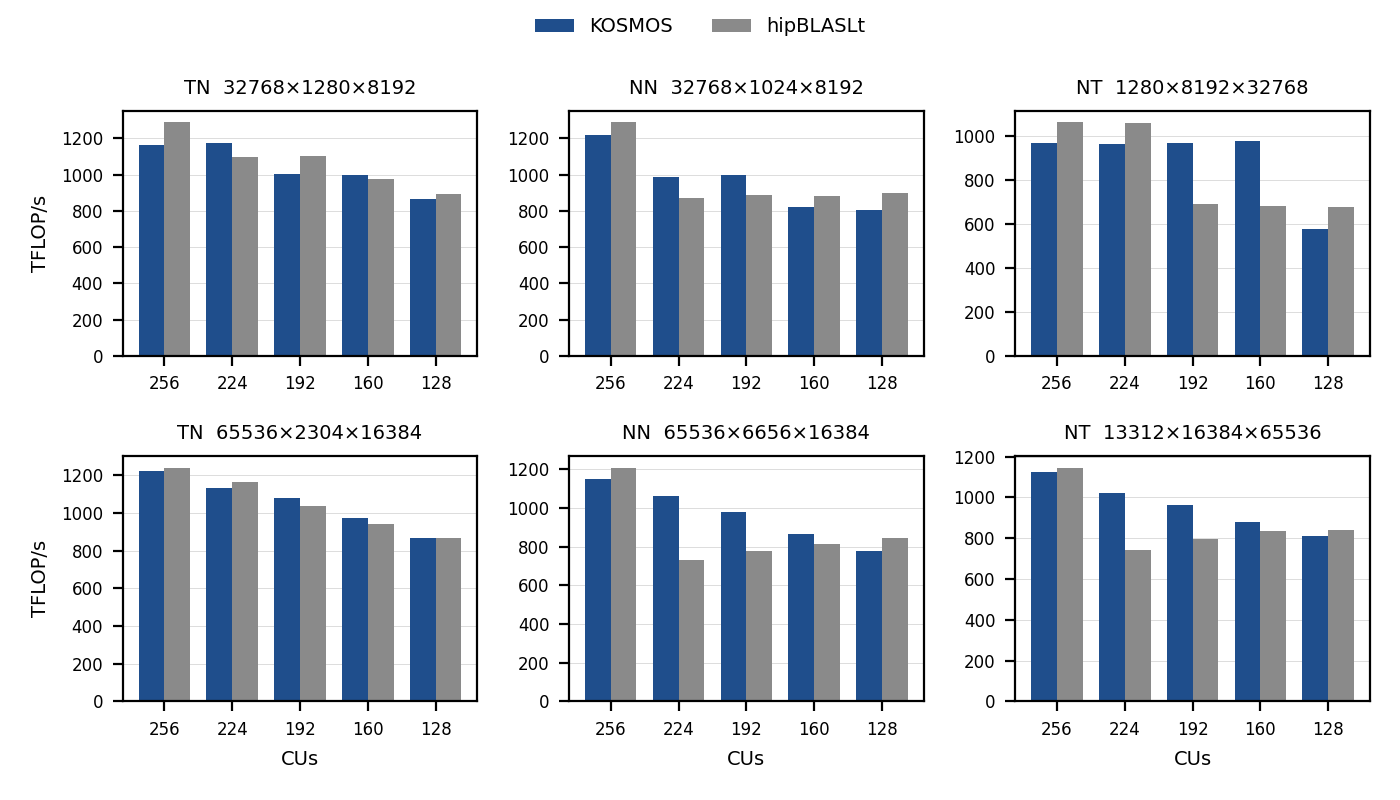}%
    \caption{GEMM-only throughput (TFLOPS) of \sysname{} GEMM kernels and hipBLASLt under a CU cap, communication off. Top: small shapes (Llama~3 70B, MBS~4); bottom: large shapes (Llama~3 405B, MBS~8); TN, NN, and NT layouts, titled M$\times$N$\times$K. Measured on MI350X.}
    \label{fig:gemmcap}
\vspace{1.5em}
    \includegraphics[width=\linewidth,height=0.40\textheight,keepaspectratio]{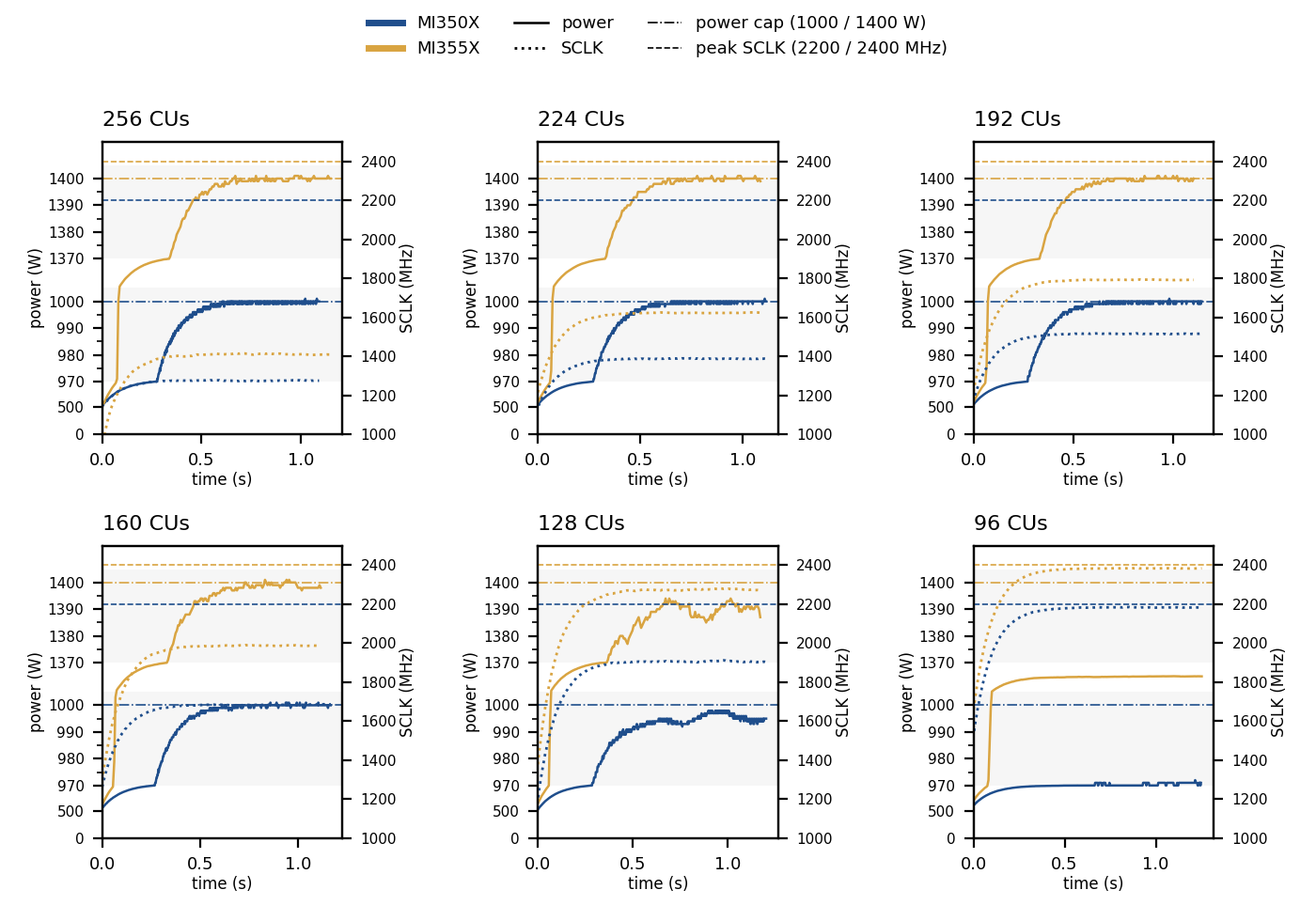}%
    \caption{Power and shader clock (SCLK) values over a Llama~3 70B-FC1-MBS8 GEMM stress test of about 1.1--1.3\,s. Both MI350X and MI355X remain at their  1000W and 1400W power caps until dropping to 128 CUs. Both GPUs reach their SCLK peak around 96 CUs.}
    \label{fig:clock}
\end{figure*}

In LLM training with power-capped GPUs, GEMM throughput decreases only slightly when CUs are capped. Both MI350X and MI355X~\cite{amd_mi350x,amd_mi355x} run at their power limit of 1000W and 1400W, respectively, with all 256 CUs active, so when a GEMM runs on fewer CUs, the remaining ones clock higher (Figure~\ref{fig:clock}). On MI350X, our GEMMs keep an average 93\% of their full-GPU throughput on 224 CUs (87.5\% of the GPU), and 88\% on 192 CUs (75\% of the GPU) (Figure~\ref{fig:gemmcap}). Reserving CUs for communication, therefore, costs the GEMM less than the CU count suggests. \textbf{R2: Use the power headroom:} Collectives are power cheap, so CUs should be allocated to communication up to the point where the transport saturates.

\begin{figure}
    \centering
    \includegraphics[width=1.0\linewidth]{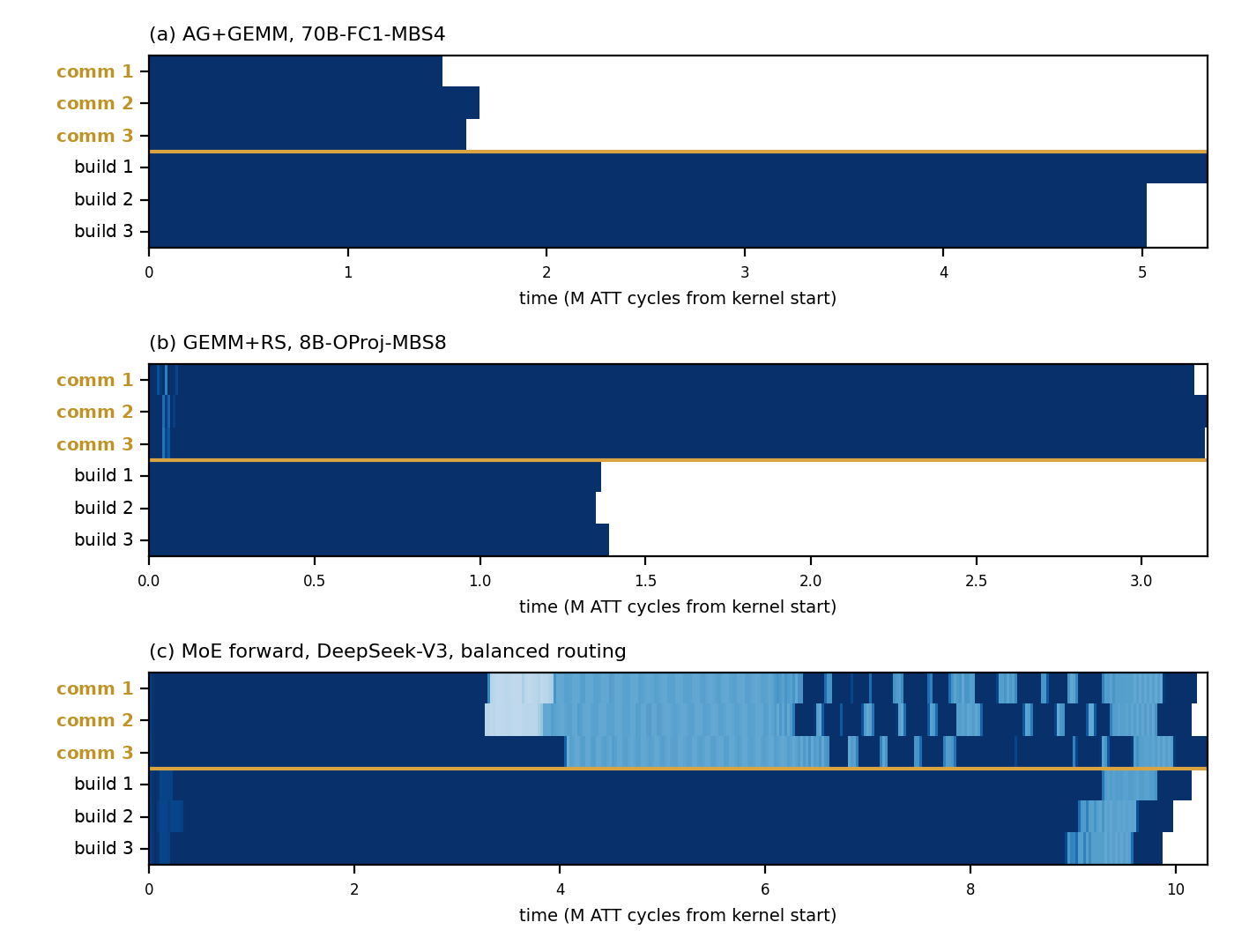}%
    \caption{Advanced Thread Trace (ATT) timeline of three communication and three computation WGs for an AG+GEMM from Llama 70B FC1, with a microbatch size (MBS) of 4; a GEMM+RS from Llama 8B with MBS 8; and MoE forward for DeepSeek-V3 with balanced routing. \emph{Dark Blue:} working. \emph{Light Blue:} spinning on a flag or sleeping. \emph{White:} exited.}
    \label{fig:trace}
\end{figure}

Overlapping a collective with its GEMM inside one kernel means dividing the GPU's CUs between communication and computation. Most fused kernels make that division once, leading to windows where WGs sit idle on CUs. For AG, this occurs when the gather has finished before compute is complete. On RS, we have two windows: (1) where communicators wait for the first data to send, and (2) at the end when compute WGs have finished their tiles but the reductions are still occurring. In A2A, dispatches act much like AG with front-loaded communication, but combine traffic is interleaved with each batched GEMM, leaving windows scattered throughout the kernel. Figure~\ref{fig:trace} shows these windows in traces of fused kernels that keep a fixed split of communication WGs. Retuning the split may reduce idle time, but does not solve this structural problem, as communication demand changes as the fused kernel progresses. \textbf{R3: Communicators budgeted adaptively:} A fixed mover count leaves either matrix cores or interconnects idle, depending on the phase of the kernel. The number of active movers should follow the work that is available.

MoE layers are a chain of steps whose sizes change from iteration to iteration. Executed as separate kernels, every step requires host overhead, routing decisions must leave the device, and no step can overlap another. Additionally, WGs do not necessarily start in their ID order (Appendix Figure~\ref{fig:wgorder}). With cross-WG dependencies (R1, R3), a resident consumer can wait on a producer that has not started; if waiting WGs fill every CU, the producer never spawns and the kernel deadlocks.  \textbf{R4: Persistent and resident:} Kernels must be persistent with all WGs resident. When CUs are needed for a second stream of work, kernels must adapt to a WG cap so the whole grid stays resident.

One key corollary follows. \textbf{R5: Self-scheduling:} While kernel design may guide WGs towards certain work, kernels cannot rely on a host round trip (breaks persistence, R4), a separate scheduling kernel (a second kernel, R1), or a scheduler WG (a CU lost to communication or compute, R2). Each WG must decide its own work only from the state it observes (adapting as demand changes, R3).

\section{Overview of \sysname{}}
\subsection{\sysname{} Design}
\sysname{} implements each fused operation as a single persistent kernel that launches one WG per CU, with each WG resident for the kernel's lifetime (R1, R4). Each WG starts in one of two roles: A \textbf{mover} transports data to and from remote peers, and a \textbf{builder} computes GEMM tiles. Depending on the collective, either role can be the producer or the consumer of the other. WGs coordinate only through in-kernel queues and readiness flags, without the host or a scheduler (R5). Rather than launching more WGs as work appears, \sysname{} keeps this fixed resident set and lets WGs change roles as the work changes (R3). Role switching is \emph{one-way} or \emph{two-way}. In one-way switching, a WG takes on another role for the rest of the kernel once its own work has finished. In AG and MoE dispatch, the movers become builders after their transfers are complete. In GEMM+RS, builders join the reduce once the GEMM tiles run out (Figure~\ref{fig:cu_timeline}(A)). In two-way switching, a WG with no work available in its current role works in the other role \emph{temporarily}, and returns when its own work appears: In the MoE combine, a set of WGs push any combine unit that is ready and builds GEMM tiles otherwise (Figure~\ref{fig:cu_timeline}(B)).

\subsubsection{Readiness and Queue Order}
Without a scheduler, two things decide how a WG finds work: the unit at which data is declared ready, and the order of the queues it claims from. A WG claims a tile by atomically taking the next entry from a shared queue; it then computes that tile. \sysname{} operates in bands, chunks, tiles, and steps. A tile row is a set of tiles spanning 256 rows. A rank's chunk is the tokens/TP-size output rows it owns after RS. A band is a group of tile rows within a chunk, and each step produces every chunk's next band.

Work is tracked at the finest grain the GEMM allows. In gathering collectives, the unit is a 256-row tile row of gathered input. In reducing collectives, it is a band of rows that is sized to the output. Every output tile has exactly one completion point, so a consumer can start work on it as soon as it is finished and enqueued.

Two rules fix the order of work: \textbf{Upstream before downstream:} A WG claims every tile of an earlier step before any tile of a later one. A WG waits only on a tile claimed earlier in queue order, to avoid deadlocks or stalls. For example, in MoE forward, GEMM2 tiles are claimed only once every GEMM1 tile has been claimed (but not necessarily finished), so a waiting tile only ever depends on the work already claimed by a resident WG (Figure~\ref{fig:cu_timeline}(B)). Additionally, a two-way WG never blocks on a push (as it builds instead), so waits are always satisfied. \textbf{Produce in the order consumers need:} If consumers are left waiting, we lose overlap. In AG+GEMM, builders work on the local shard before claiming other work while movers gather the first shards. In GEMM+RS, builders must produce band-major; finishing the first band of every rank's chunk first allows every rank's reducers to start after one step. In MoE forward, dispatch pushes experts in the order GEMM1 consumes them, and GEMM2 works on a row block so the combine can push to its destination as soon as GEMM2 finishes it.

\subsubsection{Coexistence under a CU Cap}
When other work must run concurrently, such as internode FSDP, \sysname{} kernels can be CU capped by calling frameworks to allocate CUs for a second stream of work. Then, AG+GEMM and MoE kernels shrink the builder count and RS kernels shrink both roles proportionally. Every WG of the capped grid then remains resident (R4).

\begin{figure*}[p]
\centering
\includegraphics[width=0.70\linewidth]{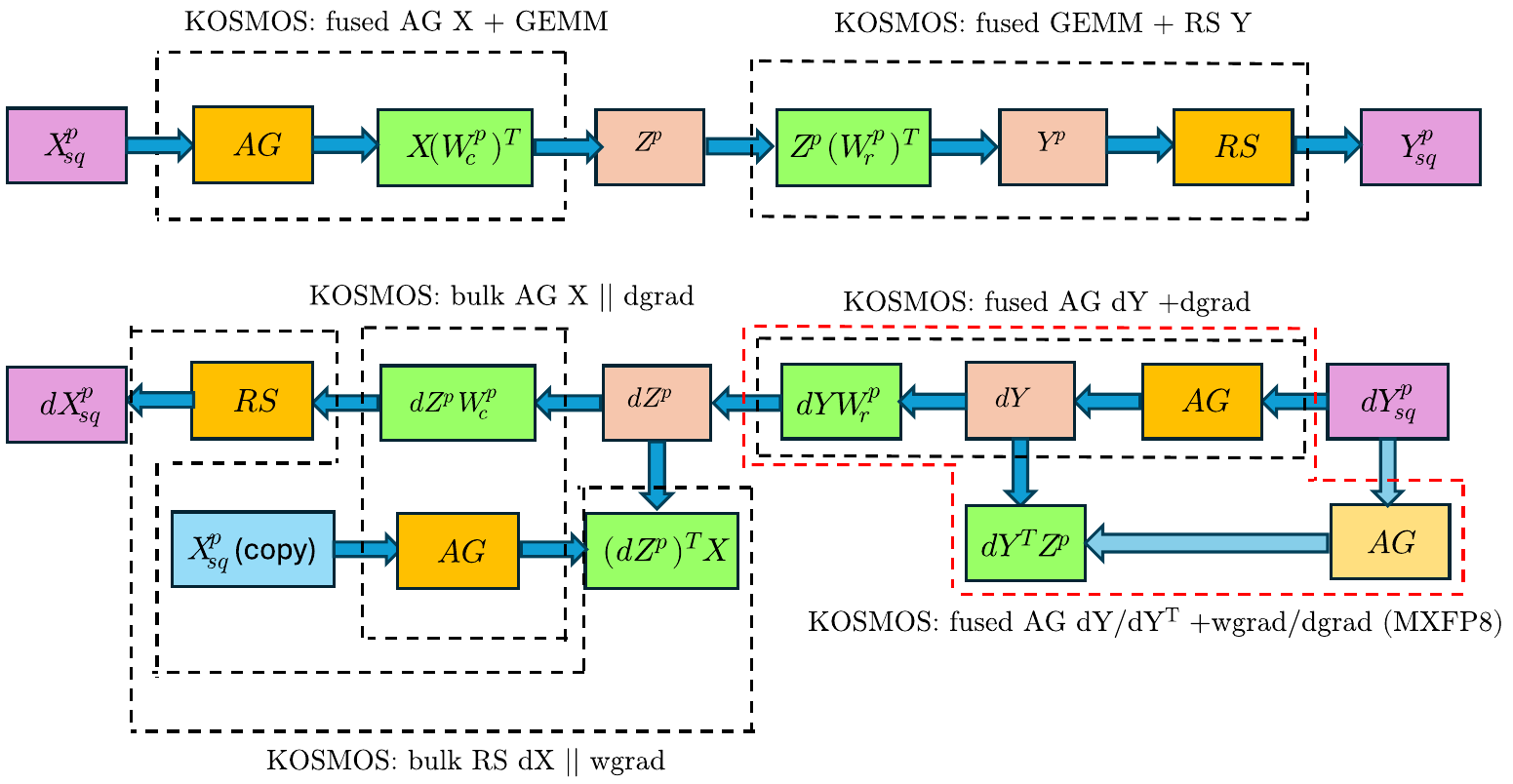}%
\caption{Overview of TP+SP kernel fusion in \sysname{}. Top: forward. Bottom: backward. Each dashed box is one \sysname{} kernel in BF16 or MXFP8; the red box is used only for MXFP8.}
\label{fig:tpsp_flowchart}
\vspace{1em}
\includegraphics[width=0.85\linewidth]{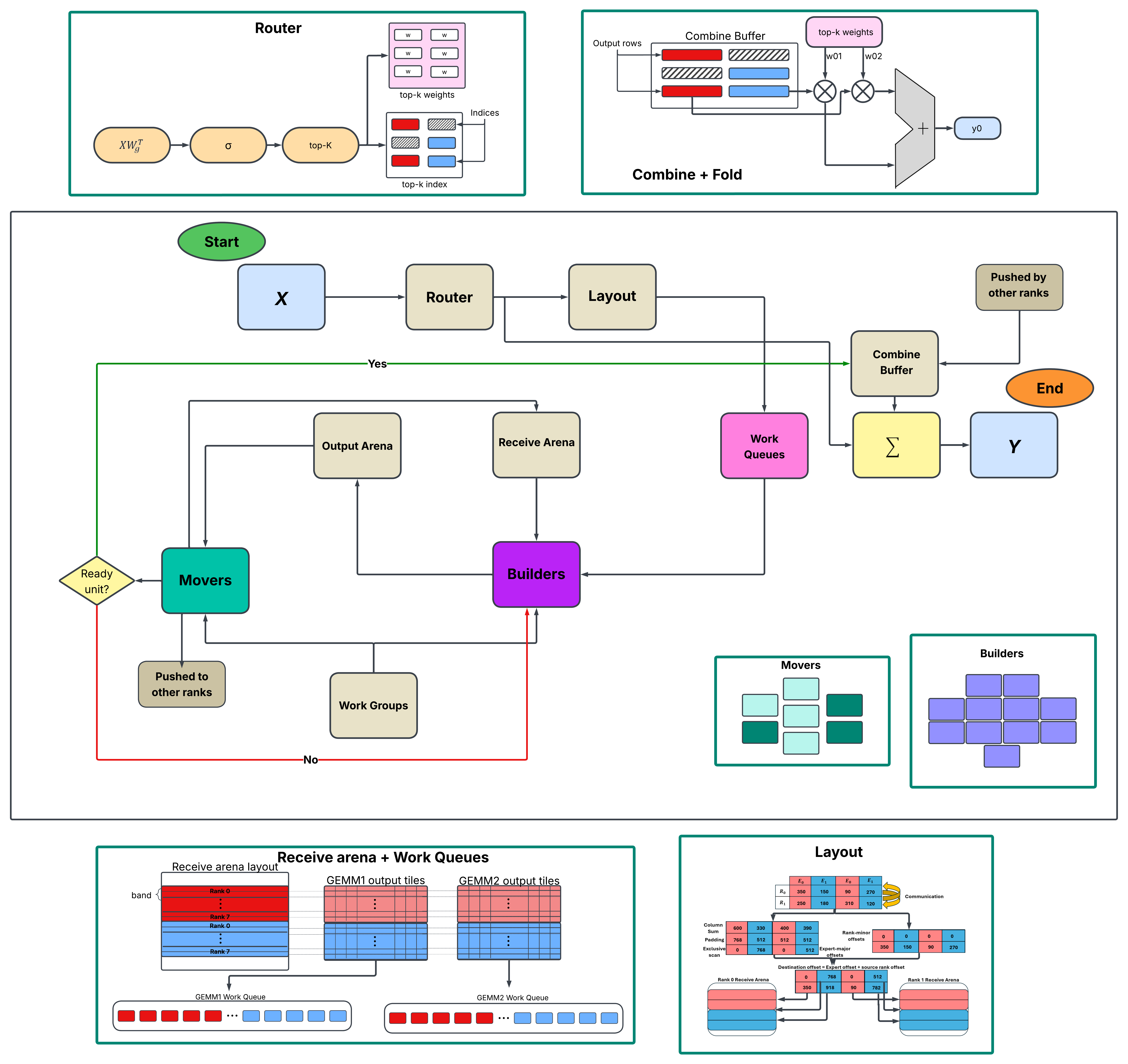}%
\caption{The \sysname{} single-launch MoE forward kernel. Red and blue: local experts; grey hatching: remote experts. Green: movers (light: dispatchers, dark: combiners); purple: builders. The flow runs from Start to End; stages are the forward-pass steps of Section~\ref{sec:moe_fwd}.}
\label{fig:moe_fwd_flow}
\end{figure*}

\subsection{TP+SP Kernels}
As shown in Figure~\ref{fig:tpsp_flowchart}, TP+SP requires six GEMMs and five or six collectives. \sysname{} names GEMMs by their layout in BLAS col-major convention: ``T'' for ``transpose,'' ``N'' for ``no transpose.'' For BF16, row-parallel wgrad is able to reuse the gathered $dY$ from row-parallel dgrad. In MXFP8 training, $dY^{T}$ is also gathered. To overlap the increased communication, both AGs are fused along with the pair of backward GEMMs for the row-parallel layer into a single kernel. Thus, \sysname{} supports five persistent kernels for both data types (Figure~\ref{fig:tpsp_flowchart}).

\subsubsection{Fused AG}
In the AG kernels (forward AG+TN, backward AG+NN), the movers pull the gathered input directly. Each mover is pinned by its WG index to one slice of one source GPU shard, and copies it one 256-row tile row at a time from the peer into a local buffer. The mover then performs a release-add on a local arrival counter for that peer and tile row. Builders compute the local shard first, then claim peer tiles in gather order. A builder spins on the counters in gather order and performs a single acquire once the count is complete. Once their slice has been transported, the movers perform a one-way switch to the builders.

In MXFP8, the row-parallel backward fuses one more GEMM: Training accuracy requires the weight gradient to use $dY$ quantized along the token dimension, so a second column-wise quantized copy of $dY$ must also be gathered. \sysname{} gathers both quantized copies of $dY$, and computes the data gradient (NN) and the weight gradient (NT) in one kernel to minimize idle time.

\subsubsection{Fused RS}
The RS kernel (forward TN+RS) uses a ``one-shot'' pull, derived from the two-shot all-reduce~\cite{doi:10.1177/1094342005051521}. Builders write their partial tiles into a local staging buffer in band-major order. Movers own a band of the output; they pull it from every rank, sum in FP32 in place in a fixed order, and round once to BF16 before storing directly to the output. Every partial crosses the fabric once and is rounded once, which has less accumulation error than RCCL's ring, and is bitwise reproducible (Appendix Table~\ref{tab:rserr}). Arrival is detected in the data itself: The staging buffer is filled with a NaN sentinel, so the landing of the data is the signal in and of itself. A BF16 signal NaN is used to avoid hangs that could arise from a matrix fused multiply-add-produced arithmetic NaN. The first builders (up to saturation) that find no GEMM tile left become movers to assist with the reduction (one-way switching).

\subsubsection{Bulk Kernels}
In the backward pass, two collectives have no data dependency on the GEMM they run beside. The AG of $X$ is needed only by the weight gradient, so it can run alongside the data gradient GEMM, and the RS of $dX$ is not needed by the weight gradient. This becomes a resource optimization problem, so \sysname{} defines a heuristic that uses GB/s/CU and TFLOPS/CU curves to balance WGs. \sysname{} then employs one-way role-switching for movers and builders in bulk fusions to minimize any remaining idle time.

\subsection{MoE Megakernels}
An MoE layer in EP routes each token to $k$ experts. Then, tokens are dispatched to the GPUs that own those experts. The per-expert token counts change every step, so they are unknown until the layer runs. To avoid host overhead and enable overlap across steps, \sysname{} launches all of the steps below in a single kernel launch (Figure~\ref{fig:moe_fwd_flow}).

\subsubsection{Forward Pass}\label{sec:moe_fwd}
\begin{enumerate}
  \item \textbf{Route.} All WGs compute the router logits
    $X W_r^{\top}$ in FP32 tiles. The last tile to finish a row block of tokens
    routes that block: score, (group-limited) top-$k$, dispatch weights, and aux-loss inputs. The last block to be routed reduces the aux loss. The routing
    never leaves the device and does not need a separate phase.
  \item \textbf{Plan.} Builders sort the routing into per-expert send lists with a stable counting sort, keeping $dW$ bitwise reproducible, since row order sets the wgrad's accumulation order and MXFP8 scale groups. Only per-expert counts cross ranks. Each rank waits for all peers' counts, then builds the layout and tile queues on the device. This wait is the forward's only cross-rank barrier.
  \item \textbf{Dispatch $\parallel$ GEMM1.} A crew of 16-96 movers (Appendix Table~\ref{tab:heur}) pushes token rows directly into their final, expert-grouped position in each owner's receive arena, which at EP~8 is slightly faster than pulling (Appendix Table~\ref{tab:pushpull}). Push lists are destination-minor, so consecutive movers target different GPUs to keep every link busy. Per-(source, expert) flags at the destination track readiness. In MXFP8, mover quantizes their rows and pad each group to 32 rows before pushing. Each GEMM1 tile, with SwiGLU fused into its epilogue, waits only for the source whose rows it reads, and stores the pre-activations for the backward. In MXFP8, an epilogue quantizes the SwiGLU output row-wise for GEMM2 and column-wise for wgrad2.
  \item \textbf{GEMM2 $\parallel$ combine.} A GEMM2 tile starts as soon as GEMM1 has
    finished its row block. Up to 128 dual-role movers (Appendix Table~\ref{tab:heur}) push each (expert, destination) unit
    back to its token owner once the unit's row blocks are done, and claim GEMM
    tiles between pushes (two-way switching). The combine always pushes in BF16, even in MXFP8 training.
  \item \textbf{Fold.} Once every row owed to this rank has arrived, each token's
    $k$ rows are weighted by its FP32 router weights and summed in FP32.
\end{enumerate}

\subsubsection{Backward Pass}
\begin{enumerate}
  \item \textbf{Dispatch.} A crew of 32-64 movers (Appendix Table~\ref{tab:heur}) pushes the router-weighted $dY$ to the expert owners. In MXFP8, $dY$ is pushed in BF16 and quantized on the receiver in both the row-wise (dgrad2) and column-wise (wgrad2) layouts. wgrad1 also needs the column-wise $X$, which the movers push after $dY$ in link time the backward leaves idle; for narrow experts where the backward is communication-bound, it is sent with the forward dispatch instead (Appendix Table~\ref{tab:heur}). After their pushes, the movers compute the routing-weight gradients $dw_j = \langle dY, Y_j\rangle$ from the forward's combine buffer and back-propagate it with the aux-loss gradient through the score function to the logit gradient.
  \item \textbf{One queue.} dgrad2, with the SwiGLU backward fused into its epilogue, is interleaved with wgrad2 of the previous expert. The router's $dX_r$ and $dW_r$ tiles are queued after dgrad2/wgrad2 to hide the tail associated with the last experts to land.
  \item \textbf{Combine $\parallel$ wgrad1.} Up to 128 dual-role movers (Appendix Table~\ref{tab:heur}) push $dX$ row blocks back to their token owners as dgrad1 completes them, while builders run wgrad1. In BF16, wgrad1 is interleaved into dgrad1; in MXFP8, wgrad1 uses the column-wise $X$, which arrives late when sent in the backward. All of dgrad1 runs first to cover its arrival; wgrad1 then overlaps the combine. 
  \item \textbf{Fold.} Each token's $k$ gradient rows and the router's $dX_r$ are
    summed in FP32 into $dX$.
\end{enumerate}

\subsection{Implementation Details}
\sysname{} is written in C++20 and HIP using eight waves and 512 threads per WG; local data share and register usage are kept within single occupancy. GEMM tiles use HipKittens~\cite{MLSYS2026_bc75fa98}, a C++ domain-specific language for AMD GPUs, with 256x256 output tiles for all expert and TP+SP GEMMs; the router GEMM uses a 64x128 tile as it only has as many columns as experts. Shape-dependent parameters (mover crew size, transport parameters, GEMM parameters) are fixed with closed-form rules during the first training iteration (Appendix Table~\ref{tab:heur}).

\section{Evaluation}
\subsection{Experimental Setup}
\subsubsection{System Specifications}
Two single-node systems were used, with eight GPUs (MI350X or MI355X~\cite{amd_mi350x,amd_mi355x}) each, fully connected over XGMI. Both nodes contained two AMD Zen 5 CPUs with 64 cores each. While MI350X runs at a peak clock of 2.2 GHz at up to 1000W, MI355X runs at a peak of 2.4 GHz at up to 1400W. MI350X offers 2.31 PFLOPS of peak dense BF16 matrix multiplication, while MI355X is slightly higher at 2.52 PFLOPS. The peak performance of MXFP8 is 4.62 PFLOPS and 5.03 PFLOPS, respectively.

The benchmarks were run on ROCm 7.14~\cite{rocm} with RCCL 2.30.4~\cite{rccl}, PyTorch~\cite{NEURIPS2019_bdbca288} 2.12, ROCm Transformer Engine~\cite{rocm_te} 2.18, and ROCm Megatron-LM~\cite{rocm_megatron} commit b0695a0.

\subsubsection{Integration}
The microbenchmarks were run with a standalone harness in \sysname{}. For TP+SP, a RCCL collective and a GEMM are run serially as a baseline (hipBLASLt~\cite{hipblaslt} in BF16, a communication-free HipKittens GEMM in MXFP8). Each configuration is run with 24 shapes arising from the Llama3~\cite{grattafiori2024llama3herdmodels} class of models. The 8B, 70B, and 405B model sizes are used with a microbatch size between 1 and 8 and sequence length~8192. MoE runs one layer, both forward and backward, in a standalone harness. Layers of nine MoE models are used: DeepSeek V3~\cite{deepseekai2025deepseekv3technicalreport}, DeepSeek V4 Pro~\cite{deepseekai2026deepseekv4highlyefficientmilliontoken}, Qwen3-30B and Qwen3-235B~\cite{yang2025qwen3technicalreport}, Qwen3.5-397B~\cite{qwen2026qwen35}, Kimi K2.7 Code~\cite{moonshot2026kimik27}, GLM 5.2~\cite{zai2026glm52}, and Llama 4 Scout and Maverick~\cite{meta2025llama4}. The MoE models use sequence length~4096 and MBS~2. \sysname{} is tested against baselines with balanced and skewed routing to simulate different training scenarios.

End-to-end runs are executed using ROCm Megatron-LM. \sysname{} TP+SP kernels are transplanted into TE and run against Megatron-LM using the default ROCm TE path. The 405B model is restricted to 24 layers to fit on a single node. For MoE, the \sysname{} adapter replaces Megatron's router, token dispatcher, grouped multilayer perceptron (MLP), and shared expert if present. Due to resource constraints, a proxy Qwen-235B model with 24 layers and a DeepSeek-V3 proxy with no dense layers and nine MoE layers are considered. TP+SP models use a global batch size (GBS) of 64, and MoE models use a GBS of 16.

\subsubsection{Numerics \& Validation}
\sysname{} kernels are bitwise run-to-run deterministic: tile claiming changes which WG computes a tile, never the reduction or fold order. Outputs are validated against RCCL+hipBLASLt with ROCm TE's tolerances~\cite{rocm_te}, re-poisoned every iteration, and checked twice to catch races. Microbenchmarks run and discard 10 warm-up iterations before 25 timed iterations. Each is run three times with the variant order shuffled, and we report the median of the slowest rank; the median run-to-run spread is 0.4\% (max 2.0\%). End-to-end results report Megatron-LM's printed iteration time and throughput. Each configuration is run three times for 20 iterations, with the median result presented. MoE TFLOPS count actual routed FLOPs.

\subsection{Ablations}\label{sec:ablations}
\subsubsection{Data Readiness Mechanisms}

Table~\ref{tab:mechanisms} isolates the design decisions behind \sysname{}, changing one mechanism at a time with everything else held fixed. The largest effects come from the order in which data is produced and consumed: local-first AG queues consumed in gather order, band-major RS production, and destination-minor A2A push lists. \sysname{} is optimized for bandwidth per CU: its 8-wave movers saturate XGMI links with half the CUs of a 4-wave kernel (Appendix Table~\ref{tab:waves}). Fusing the router and shared expert into the MoE launch is a small but consistent gain. Readiness-driven switching only pays off if readiness arrives where WGs look next: Thus, the order of production is as important as the dynamic role switching it enables.

\begin{table*}[t]
\caption{Design mechanisms ablated in isolation: slowdown of the alternative relative to \sysname{}. Full results in Appendix~\ref{app:ablations}.}
\label{tab:mechanisms}
\vskip 0.1in
\begin{center}
\begin{small}
\begin{sc}
\setlength{\tabcolsep}{3pt}
\begin{tabular}{lll}
\toprule
Mechanism & Alternative & Alternative vs \sysname{} \\
\midrule
AG queue order       & no local-first, chunk-major & 1.17--1.40$\times$ slower \\
RS production order  & row-major                   & 1.41--1.45$\times$ slower \\
A2A push order       & destination-major           & 1.12--2.03$\times$ slower \\
8-wave movers        & 4 waves per WG (7--56 WGs)  & 1.23--1.96$\times$ slower \\
Router fusion        & separate router kernels     & 1.01--1.02$\times$ slower \\
Shared-expert fusion & separate kernels            & 1.02--1.04$\times$ slower \\
\bottomrule
\end{tabular}
\end{sc}
\end{small}
\end{center}
\vskip -0.1in
\end{table*}

\subsubsection{Role-Switching Policy}

To measure the impact of \sysposs{} derived role switching policies, we compare them to several alternatives (Table~\ref{tab:policy}). For TP+SP, as implied by the one-way relationship of AG/RS and their GEMMs, we found that one-way role switching overall outperformed two-way role switching. One-way role switching is never more than 1\% slower than the best single fixed split, and up to 1.15$\times$ faster. For AG, communication-bound shapes generally do not benefit from role switching, as movers are active until the very last step: While MLP layers improve by a geometric mean of 1.05$\times$, attention layers average 1.00$\times$. Per-shape results for both GPUs are in Appendix Figures~\ref{fig:tpsppolicy} and~\ref{fig:tpsppolicy355}.

For MoE, we show the benefits of our two-way role switch in \sysname{} over alternatives. \sysname{} uses an interleaved two-way role switching policy, where WGs decide their role after every tile. We compare this to a policy where dispatch is executed with one-way role switching and combine uses a dedicated WG crew, following UniEP~\cite{10.1145/3806645.3807818}, and to a static two-way planned steal, following Weave~\cite{huang2026weavefinegraineddynamicsm}, where movers finish dispatch, compute a predetermined number of tiles from a runtime cost model, then return to combine. We find the two-way role switch not only beats a fixed split in all 36 cells on both MI350X and MI355X, but is the best policy on average, with a geometric mean speedup of 1.04$\times$/1.04$\times$ on the forward pass and 1.06$\times$/1.08$\times$ on the backward pass (MI350X/MI355X). This gain is up to 1.14$\times$/1.16$\times$. Appendix Figures~\ref{fig:moepolicy} and~\ref{fig:moepolicy355} show the speedups per shape on the MI350X and MI355X, respectively.

\begin{table*}[t]
\centering
\small
\caption{Role-policy ablation: geometric-mean speedup over a fixed communication crew. TP+SP: 24 shapes per kernel, fixed crew of 56 WGs. MoE: nine models, fixed crew is the measured peak per shape, also used by the dedicated combine crew (following UniEP~\cite{10.1145/3806645.3807818}). Per-step planned split following Weave~\cite{huang2026weavefinegraineddynamicsm}. All policies are implemented in \sysname{}. $^{\ast}$Shipped \sysname{} policy; bold: fastest per row and GPU.}
\label{tab:policy}
\vskip 0.1in
\setlength{\tabcolsep}{4pt}
\begin{tabular*}{\textwidth}{@{\extracolsep{\fill}}l cccc cccc@{}}
\toprule
 & \multicolumn{4}{c}{MI350X} & \multicolumn{4}{c}{MI355X} \\
\cmidrule(lr){2-5}\cmidrule(l){6-9}
 & One-way & Two-way & Dedicated & Planned & One-way & Two-way & Dedicated & Planned \\
 & switch & switch & combine & split & switch & switch & combine & split \\
\midrule
TP+SP AG+TN        & \textbf{1.032}$^{\ast}$ & 0.965 & -- & -- & \textbf{1.033}$^{\ast}$ & 0.932 & -- & -- \\
TP+SP TN+RS        & \textbf{1.031}$^{\ast}$ & 1.024 & -- & -- & \textbf{1.018}$^{\ast}$ & 1.012 & -- & -- \\
TP+SP AG+NN        & \textbf{1.018}$^{\ast}$ & 0.951 & -- & -- & \textbf{1.017}$^{\ast}$ & 0.918 & -- & -- \\
\midrule
MoE fwd, balanced  & -- & \textbf{1.029}$^{\ast}$ & 1.018 & 0.985 & -- & \textbf{1.027}$^{\ast}$ & 1.008 & 0.982 \\
MoE fwd, skewed    & -- & \textbf{1.047}$^{\ast}$ & 1.029 & 0.989 & -- & \textbf{1.051}$^{\ast}$ & 1.019 & 0.914 \\
MoE bwd, balanced  & -- & \textbf{1.052}$^{\ast}$ & 1.012 & 0.973 & -- & \textbf{1.067}$^{\ast}$ & 1.009 & 0.836 \\
MoE bwd, skewed    & -- & \textbf{1.064}$^{\ast}$ & 1.021 & 0.963 & -- & \textbf{1.093}$^{\ast}$ & 1.009 & 0.752 \\
\bottomrule
\end{tabular*}
\end{table*}

\subsection{TP+SP Speedups}

\subsubsection{TP+SP Microbenchmarks}

\begin{figure}
    \centering
    \includegraphics[width=1\linewidth]{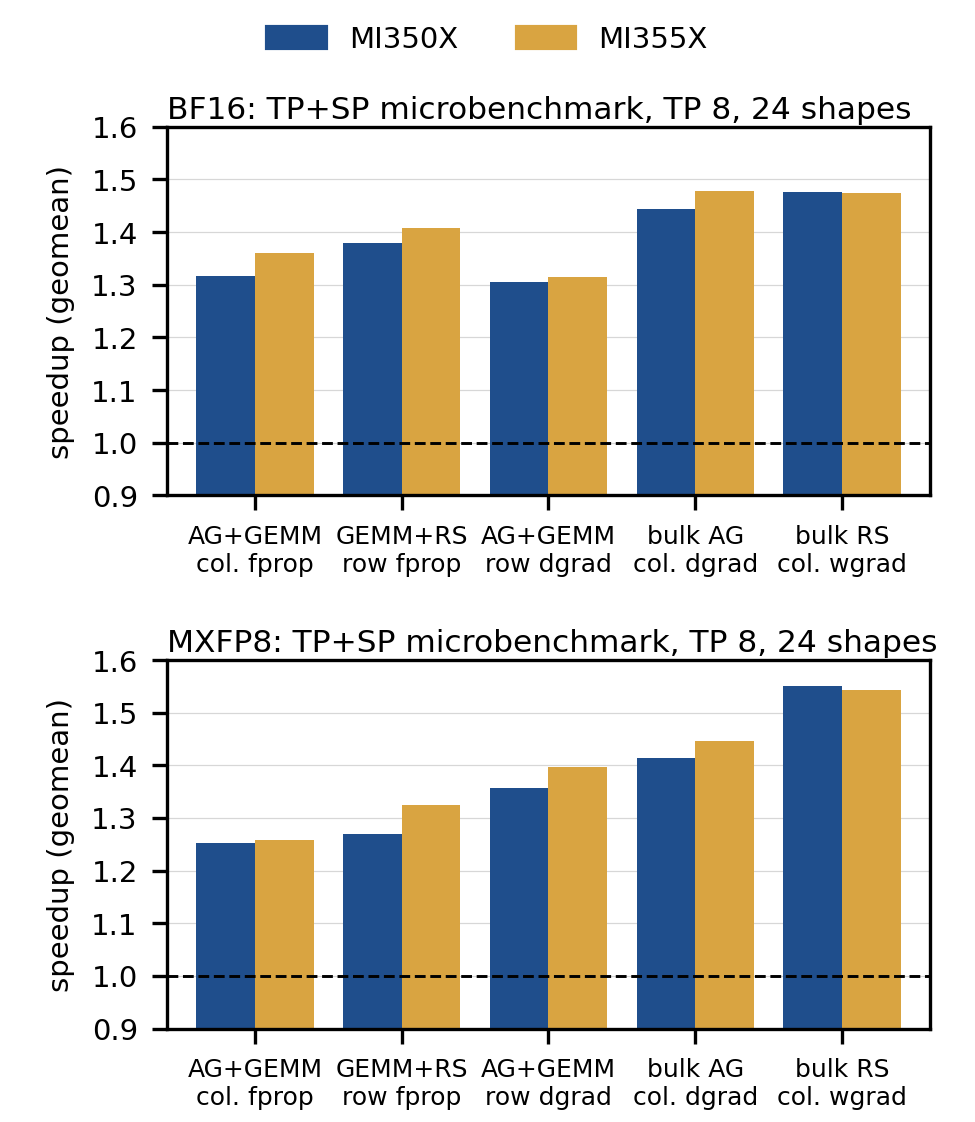}%
    \caption{TP+SP microbenchmark: \sysname{} speedup over the baseline, geometric mean over 24 Llama~3 shapes per config.}
    \label{fig:tpspmicro}
\end{figure}

Figure~\ref{fig:tpspmicro} compares each \sysname{} kernel to the same RCCL collective and hipBLASLt/HipKittens GEMM run without overlap. \sysname{} is faster on all 480 combinations of shape, kernel, precision, and GPU. In BF16, we see a geometric mean speedup of 1.38$\times$ on MI350X and 1.41$\times$ on MI355X. In MXFP8, the geometric mean is 1.36$\times$ on MI350X and 1.39$\times$ on MI355X. The MXFP8 GEMMs have twice the peak throughput of BF16, while an MXFP8 all-gather moves 33/64 of the BF16 bytes and the reduce-scatters move BF16 in both precisions, which leaves less compute to hide. Compared against unfused \sysname{} communication and GEMM kernels run without overlap gives similar geometric mean speedups: For BF16, the speedup is 1.39$\times$/1.42$\times$, and for MXFP8 the speedup is 1.37$\times$/1.40$\times$. This shows that most of the benefit over the baselines is due to the overlap mechanisms, rather than the components.

\subsubsection{TP+SP End-to-End Training}

Table~\ref{tab:e2e-tpsp-main} reports Llama~3 training in Megatron with TP~8, where \sysname{} replaces Transformer Engine's TP+SP collectives and GEMMs. \sysname{} is faster for every model size, precision, and GPU model.

\begin{table}[h!]
\caption{Llama~3 TP+SP end-to-end training, TP~8, sequence 8192, GBS~64: \sysname{} speedup over ROCm Transformer Engine (range over MBS 1--8, same MBS in both arms) and TFLOPS/GPU at each arm's best MBS (TE $\to$ \sysname{}). \textsuperscript{r}Full recompute in both arms; 70B at MBS~4 and~8 runs all 80 layers, and 405B runs 24 of its 126 layers. Per-MBS results: Appendix Table~\ref{tab:e2e-tpsp}.}
\label{tab:e2e-tpsp-main}
\vskip 0.1in
\begin{center}
\begin{footnotesize}
\setlength{\tabcolsep}{1.5pt}
\begin{tabular}{llcccc}
\toprule
 & & \multicolumn{2}{c}{MI350X} & \multicolumn{2}{c}{MI355X} \\
\cmidrule(lr){3-4}\cmidrule(lr){5-6}
Model & Prec. & Speedup & TFLOPS & Speedup & TFLOPS \\
\midrule
8B & BF16 & 1.20--1.27 & 474$\to$596 & 1.24--1.28 & 478$\to$612 \\
 & MXFP8 & 1.21--1.25 & 580$\to$718 & 1.22--1.24 & 597$\to$730 \\
\midrule
70B & BF16 & 1.15--1.21 & 664$\to$769 & 1.22--1.28 & 654$\to$834 \\
 & MXFP8 & 1.19--1.22 & 893$\to$1067 & 1.21--1.23 & 907$\to$1119 \\
\midrule
405B & BF16 & 1.10--1.16 & 570\textsuperscript{r}$\to$645\textsuperscript{r} & 1.17--1.22 & 624\textsuperscript{r}$\to$757\textsuperscript{r} \\
 & MXFP8 & 1.10--1.14 & 862\textsuperscript{r}$\to$985\textsuperscript{r} & 1.16--1.18 & 947\textsuperscript{r}$\to$1101\textsuperscript{r} \\
\bottomrule
\end{tabular}
\end{footnotesize}
\end{center}
\vskip -0.1in
\end{table}

\subsection{Mixture-of-Experts Speedups}

\subsubsection{MoE Microbenchmarks}

\begin{figure}[t]
    \centering
    \includegraphics[width=1\linewidth]{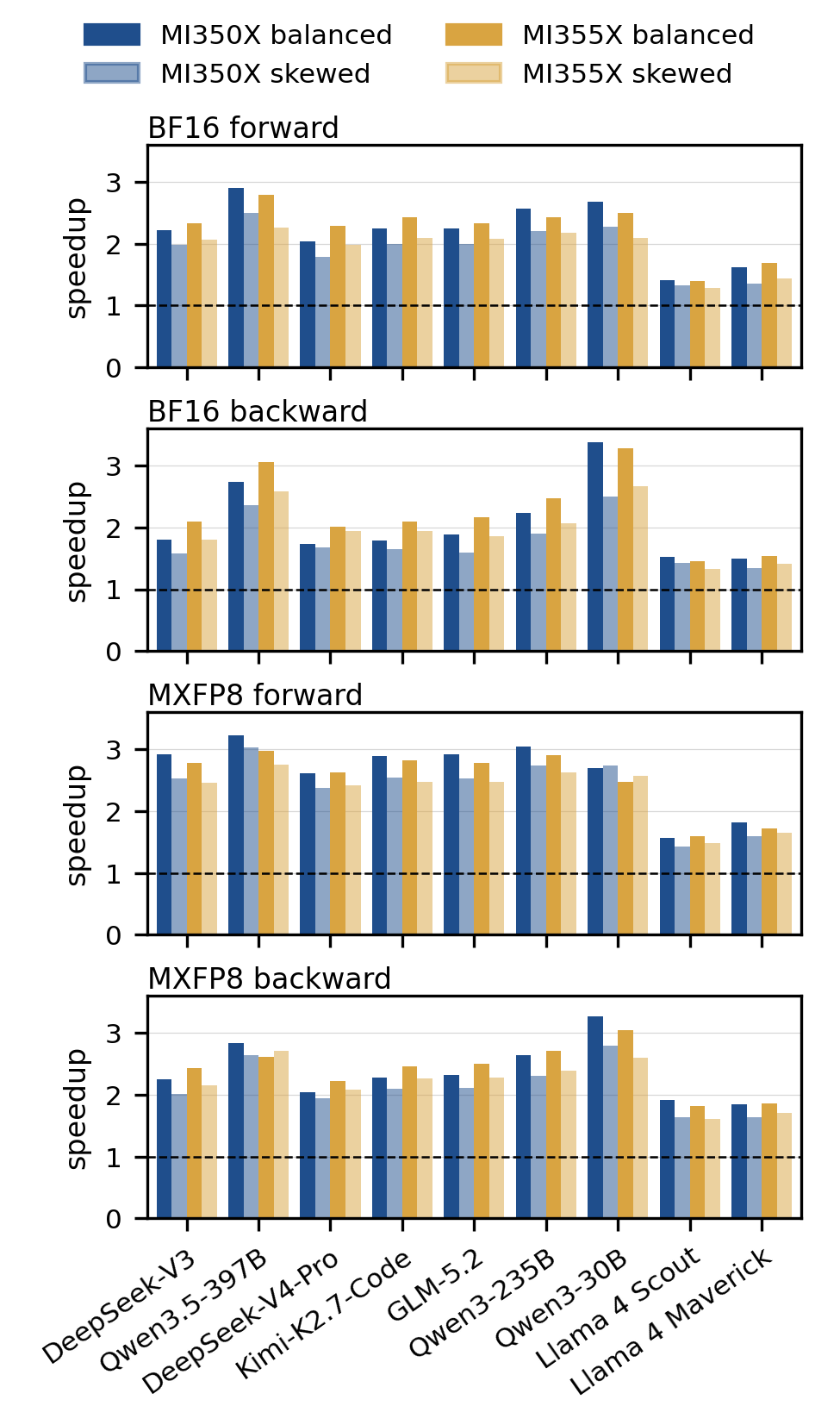}%
    \caption{MoE layer speedups over baselines for \sysname{}. Bars are per-model speedup for each direction, precision, and routing.}
    \label{fig:moemicro}
\end{figure}

Figure~\ref{fig:moemicro} compares one \sysname{} MoE layer against RCCL A2A with TE's grouped GEMMs. The figure covers nine models, both forward and backward, with balanced and skewed routing. \sysname{} is faster in all 144 cases, 1.28--3.39$\times$. Gains shrink as GEMM compute increases and top-$k$ decreases: Qwen3-30B (narrow experts, top-8) gains up to 3.39$\times$, while Llama~4 (wide experts, top-1) gains 1.28--1.92$\times$. The forward pass with balanced routing gains the most, and backward pass with skewed routing is sped up the least. We also run \sysname{} components as separate launches; they are 1.25--1.51$\times$ slower than the fused layer on geometric mean which we attribute to kernel fusion.

\subsubsection{MoE Proxy Training}
Table~\ref{tab:e2e-moe-main} reports end-to-end training of two MoE proxy models. For both, balanced refers to random routing with a uniform distribution, so each expert has an equal expected number of tokens. Skewed results refer to a pure top-k routing defined by each model. For DeepSeek V3 with Megatron-generated mock data, this was seen to be up to 1.3$\times$ the data of balanced for the largest peer, whereas Qwen saw skew up to 4.2$\times$ in some iterations. \sysname{} relies on Megatron's capacity factor flag to size shared buffers accordingly. A greater skew falls back to the default Megatron MoE layer. \sysname{} is faster than Megatron-LM with ROCm's DeepEP~\cite{deepep2025,rocm} in every configuration, by 1.29--1.83$\times$ across GPU models.

\begin{table}[h!]
\caption{MoE proxy training, EP~8: \sysname{} speedup over DeepEP, and TFLOPS/GPU (DeepEP $\to$ \sysname{}). Balanced: uniform random routing; skewed: the model's top-$k$ router. Full results: Appendix Table~\ref{tab:e2e-moe}.}
\label{tab:e2e-moe-main}
\vskip 0.1in
\begin{center}
\begin{footnotesize}
\setlength{\tabcolsep}{2pt}
\begin{tabular}{llcccc}
\toprule
 & & \multicolumn{2}{c}{MI350X} & \multicolumn{2}{c}{MI355X} \\
\cmidrule(lr){3-4}\cmidrule(lr){5-6}
Routing & Prec. & Speedup & TFLOPS & Speedup & TFLOPS \\
\midrule
\multicolumn{6}{l}{\textbf{Qwen3-235B}} \\
balanced & BF16 & 1.70 & 359$\to$610 & 1.74 & 387$\to$673 \\
 & MXFP8 & 1.82 & 372$\to$677 & 1.83 & 400$\to$731 \\
skewed & BF16 & 1.31 & 249$\to$327 & 1.29 & 275$\to$355 \\
 & MXFP8 & 1.57 & 245$\to$384 & 1.54 & 268$\to$411 \\
\midrule
\multicolumn{6}{l}{\textbf{DeepSeek-V3}} \\
balanced & BF16 & 1.31 & 423$\to$556 & 1.33 & 480$\to$637 \\
 & MXFP8 & 1.39 & 446$\to$619 & 1.39 & 500$\to$696 \\
skewed & BF16 & 1.33 & 414$\to$551 & 1.36 & 465$\to$631 \\
 & MXFP8 & 1.40 & 437$\to$613 & 1.41 & 488$\to$689 \\
\bottomrule
\end{tabular}
\end{footnotesize}
\end{center}
\vskip -0.1in
\end{table}

\section{Related Work}\label{sec:related}
\textbf{Stream-level overlap.} Frameworks~\cite{li2020pytorchdistributedexperiencesaccelerating,nvidia_te} run NCCL/RCCL~\cite{nccl,rccl} collectives on separate streams, where they hold CUs for the whole transfer and compete with the GEMM. CoCoNet~\cite{10.1145/3503222.3507778} adds device-side locks, chunking~\cite{10.1145/3567955.3567959,ICLR2025_e6231c5f} shrinks the GEMMs, and copy engines~\cite{zheng2025tritondistributedprogrammingoverlappingkernels} or new hardware~\cite{10.1145/3620665.3640410} cannot reduce or signal per tile.

\textbf{Fused communication kernels.} FLUX~\cite{chang2024fluxfastsoftwarebasedcommunication} fuses RS into a GEMM epilogue and AG into its prologue; ParallelKittens~\cite{MLSYS2026_ff997469}, TileLink~\cite{MLSYS2025_c6ee784c}, and Iris~\cite{awad2025irisfirstclassmultigpuprogramming} fuse TP+SP collectives per tile with a tuned number of communication streaming multiprocessors (SMs) or WGs. For MoE, COMET~\cite{MLSYS2025_e27ea0cd}, Mixture-of-Kittens~\cite{sul2026mixtureofkittensmoemegakernelnvl72s}, and the MegaMoE kernels of DeepSeek-V4~\cite{deepgemm2025,deepseekai2026deepseekv4highlyefficientmilliontoken} and Primus-Turbo~\cite{primusturbo_megamoe,zhang2026enabling} fix communication resources at build time or per shape, before routing is known; Mixture-of-Kittens (a second backward megakernel) and Primus-Turbo (fused operators reused in backward) compute the router outside the kernel. FlashMoE~\cite{NEURIPS2025_918d938b} and MPK~\cite{318585} are persistent inference kernels driven by scheduler warps. FLUX, ParallelKittens, COMET, and Mixture-of-Kittens are NVIDIA-only, so Section~\ref{sec:ablations} instead reproduces the UniEP and Weave policies inside the \sysname{} kernels.

\textbf{Adaptive communication budgets.} UniEP~\cite{10.1145/3806645.3807818} orders one MoE task queue communication-first, with a split set by a cost model and tuner; Weave~\cite{huang2026weavefinegraineddynamicsm} sizes its communication SMs with an in-kernel cost model and lets them steal a precomputed number of GEMM tiles; mKernel~\cite{mao2026mkernelfastmultigpumultinode} adapts the split with an in-kernel controller but trails its best fixed split. In \sysname{}, roles follow data readiness alone, with no scheduler, task graph, per-call cost model, or steal budget, and one launch per direction covers every TP+SP collective with its GEMM, and the whole MoE layer, router and gradients included.

\section{Conclusion}
Current state-of-the-art overlaps communication in TP+SP and EP training by running it on a separate stream, or by giving it a fixed or predicted share of the GPU inside a fused kernel. We show that on CDNA4 GPUs, both approaches are sub-optimal. Concurrent streams may contend without coordination, and a fixed split strands CUs whenever the demand for communication changes. \sysname{} addresses both by running communication and computation in one self-scheduling persistent kernel per operation or layer. Its WGs switch between moving and building as the work available changes, with in-kernel queues rather than a separate scheduler. Because collectives are limited by CUs up to saturation, giving CUs to communication costs little GEMM throughput, if the kernel adapts. On MI350X and MI355X nodes, \sysname{} speeds up dense model training by up to 1.28$\times$ (BF16) and 1.25$\times$ (MXFP8), and MoE training by up to 1.74$\times$ (BF16) and 1.83$\times$ (MXFP8) over DeepEP.

\sysname{} kernels are persistent only for intranode transports: XGMI with hipIPC memory is currently required. Remote direct memory access support and likely a hierarchical set of movers would carry the design to multinode. Thus, end-to-end results are limited to a single-node proxy model. We have evaluated with quasi-FSDP and pipeline parallelism stress tests, but complete multinode verification is future work. Finally, our structure requires one completion point per output tile. This disqualifies GEMM algorithms, such as stream-K~\cite{10.1145/3572848.3577479}, that compute a single output tile using multiple WGs. Other workloads built on fine-grained collectives, such as all-reduce as used in TP-only training and inference; SUMMA~\cite{van1997summa} and sparse distributed GEMMs; 3D fast Fourier transforms and spectral methods; and stencil solvers could use the same role-switching kernel design.

Overlapping collectives with computation is no longer a question of how many CUs to allocate to communication, but of when. \sysname{} answers this with self-scheduling kernels that produce in the order consumers need, and lets data readiness drive role switches.

\label{kosmos:refs-start}
\bibliography{kosmos_arxiv}
\bibliographystyle{mlsys2025}

\clearpage

\appendix
\onecolumn
\counterwithin*{figure}{section}\counterwithin*{table}{section}
\renewcommand{\thefigure}{\thesection.\arabic{figure}}\renewcommand{\thetable}{\thesection.\arabic{table}}
\label{kosmos:appendix-start}
\section{Ablations}\label{app:ablations}
\FloatBarrier
\subsection{WGs out of Order}

While workgroups of a persistent grid are all resident, they do not start in ID order. The dispatcher has tendencies to fill sorted by XCD rather than ID, but this is not guaranteed either. With oversubscription, the order depends on per-WG work. Variable work per WG can lead to highly unpredictable start order.

\begin{figure*}[!ht]
    \centering
    \includegraphics[width=1\linewidth]{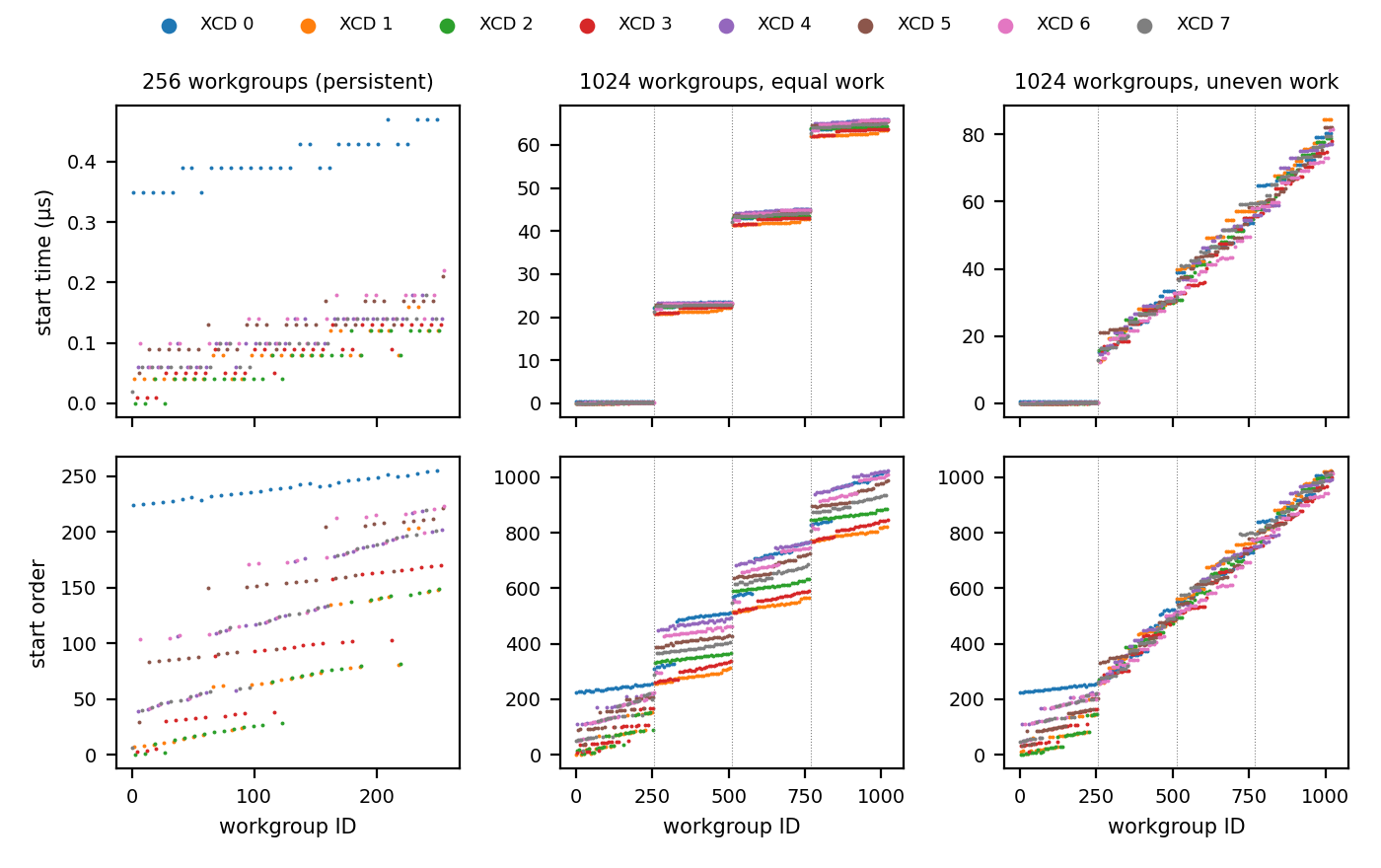}%
    \caption{Workgroup launch order and time sorted plotted against WG ID number. For persistent kernels, all WGs are resident but order is not guaranteed, and instead leans towards an accelerator complex die (XCD) ID-sorted launch order. With oversubscription, 1024 WGs exhibit different behavior based on the workload assigned to each WG.}
    \label{fig:wgorder}
\end{figure*}

\FloatBarrier
\subsection{Role Switching Policies}

These figures give the per-shape results behind Table~\ref{tab:policy}. For TP+SP, one-way switching is never significantly slower than the best fixed split, but up to 15\% faster. For MoE, two-way switching beats a fixed split in all 72 cells, and is faster than a dedicated split in 63 of 72 cells and within 2\% in the rest. Planned splits appear to lose where routing is skewed within a step, as a fixed crew size is determined before data arrives. 

\begin{figure*}[!ht]
    \centering
    \includegraphics[width=1\linewidth]{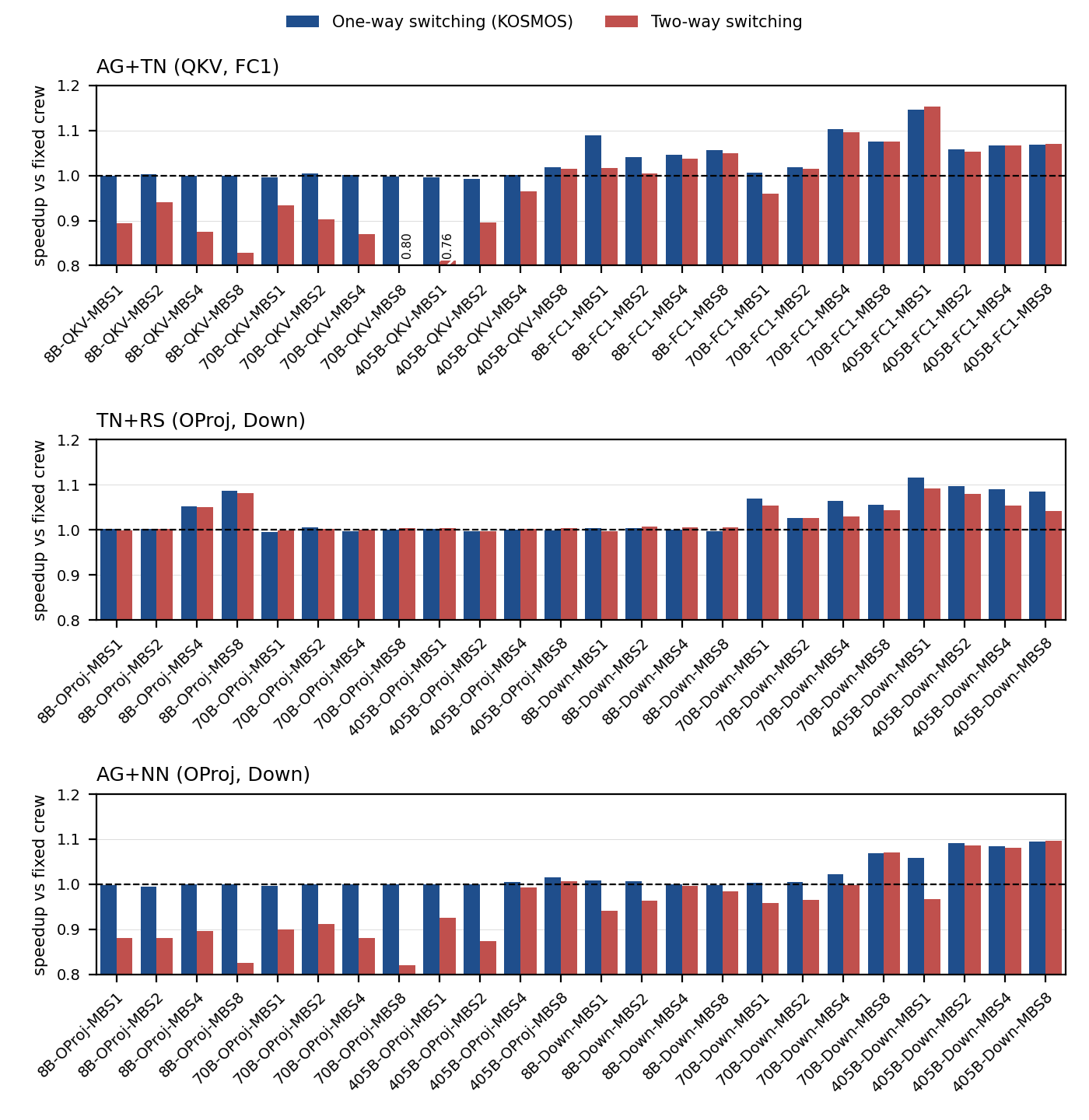}%
    \caption{MI350X: speedup of one-way and two-way role switching over a best single fixed split of 56 communication WGs (8 per peer) for the three fused formats in training: AG+TN GEMM, TN GEMM+RS, and AG+NN GEMM. Switching arms use the \sysname{} heuristic crew (Table~\ref{tab:heur}).}
    \label{fig:tpsppolicy}
\end{figure*}

\begin{figure*}[!ht]
    \centering
    \includegraphics[width=1\linewidth]{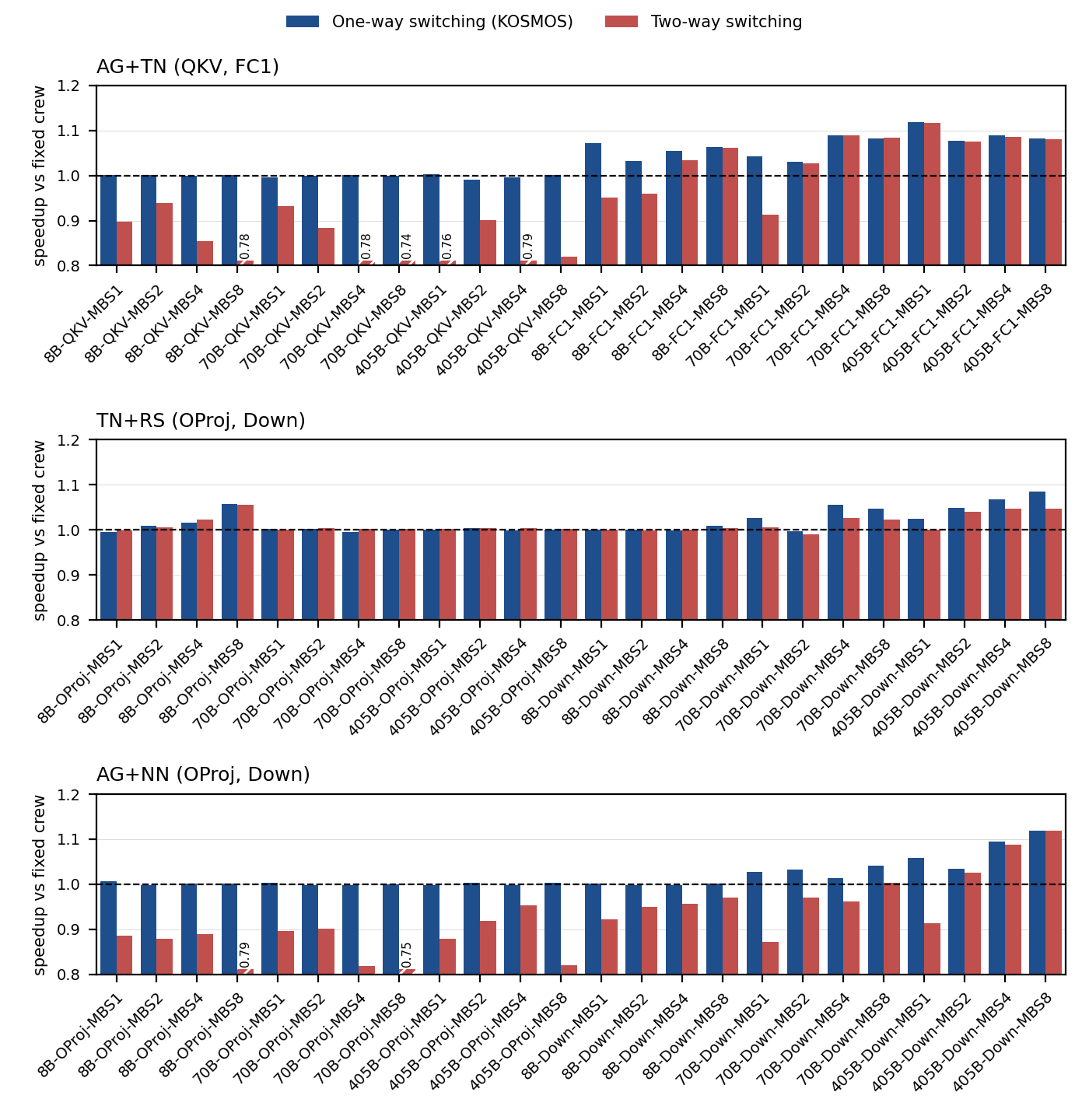}%
    \caption{MI355X: speedup of one-way and two-way role switching over a best single fixed split of 56 communication WGs (8 per peer) for the three fused formats in training: AG+TN GEMM, TN GEMM+RS, and AG+NN GEMM. Switching arms use the \sysname{} heuristic crew (Table~\ref{tab:heur}).}
    \label{fig:tpsppolicy355}
\end{figure*}

\begin{figure*}[!ht]
    \centering
    \includegraphics[width=\linewidth,height=0.38\textheight,keepaspectratio]{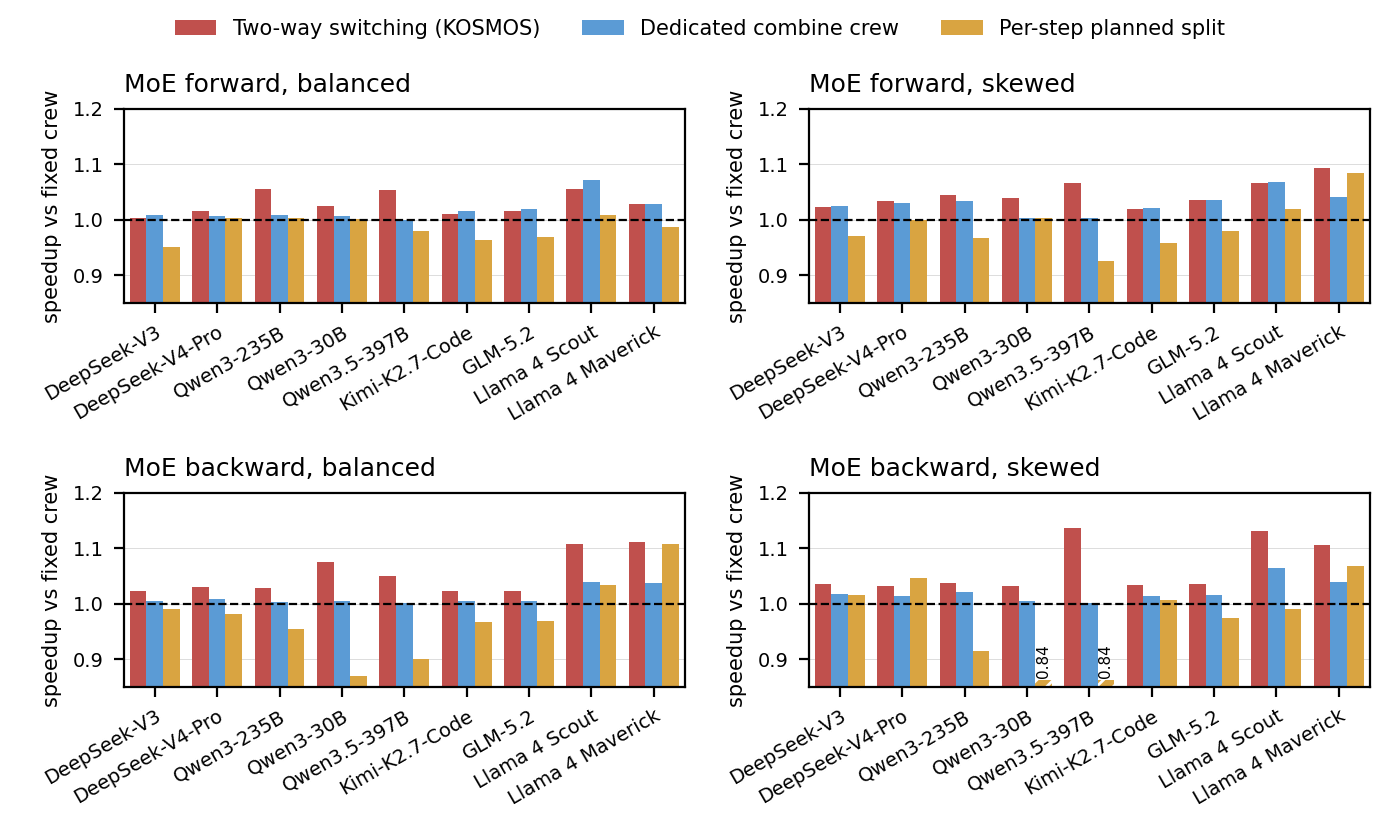}%
    \caption{MI350X: speedup over a fixed split of different role-switching policies in MoE. \sysname{} two-way role switching with data readiness role switching, a UniEP-style one-way role switching for dispatch with a dedicated combine crew, and a Weave-style planned split. Crew size for non-\sysname{} policies is the smallest that reaches 98\% of the peak dispatch and combine bandwidth (24, 32, or 48 WGs)}
    \label{fig:moepolicy}
\vspace{1.5em}
    \includegraphics[width=\linewidth,height=0.38\textheight,keepaspectratio]{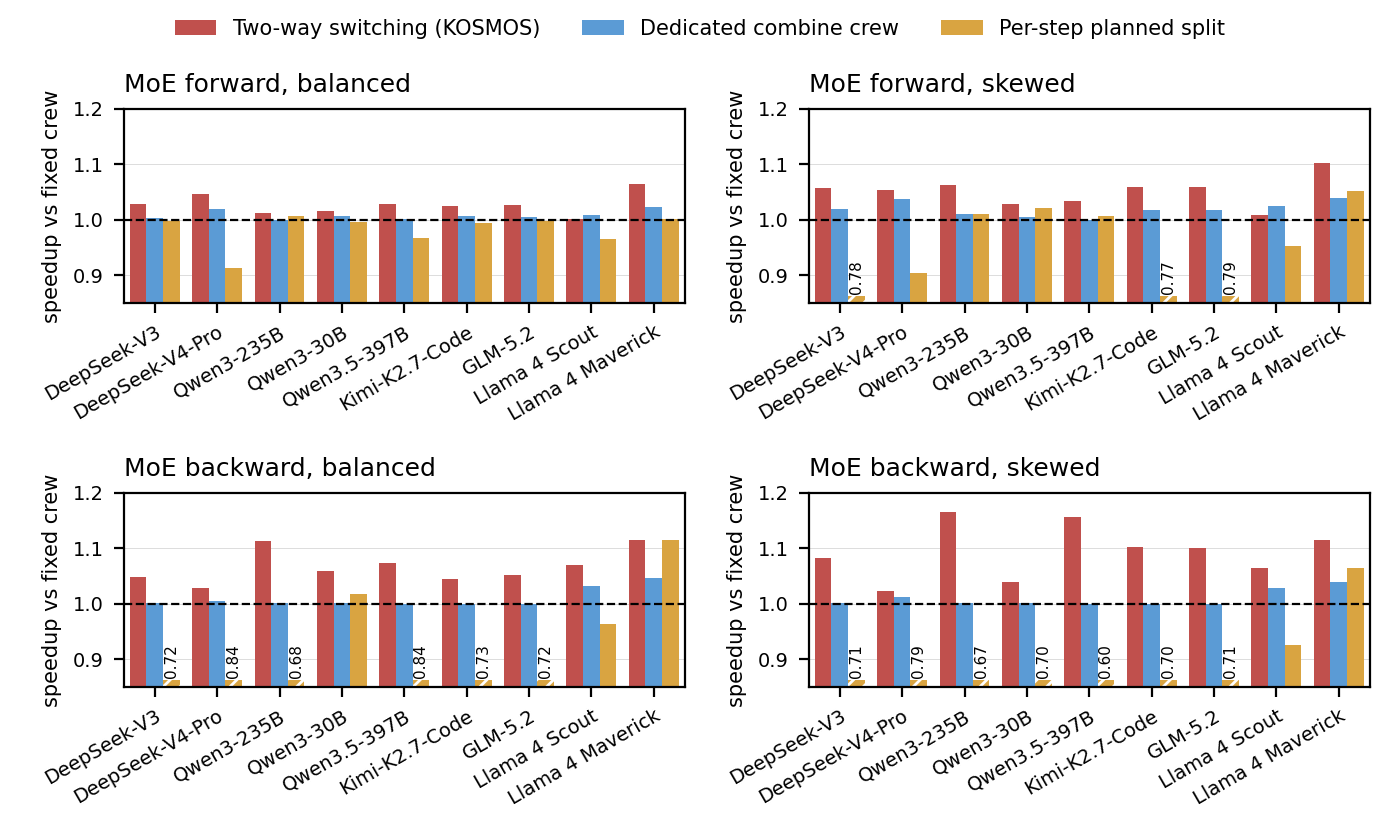}%
    \caption{MI355X: speedup over a fixed split of different role-switching policies in MoE. \sysname{} two-way role switching with data readiness role switching, a UniEP-style one-way role switching for dispatch with a dedicated combine crew, and a Weave-style planned split. Crew size for non-\sysname{} policies is the smallest that reaches 98\% of the peak dispatch and combine bandwidth (16, 24, 32, or 48 WGs)}
    \label{fig:moepolicy355}
\end{figure*}

\FloatBarrier
\subsection{Fused vs Separate Router and Shared Expert}
Running the router or the shared expert as separate kernels is 1.01--1.02$\times$ and 1.02--1.04$\times$ slower, respectively, than the fused launch (Table~\ref{tab:fusion}).

\begin{table*}[!ht]
\centering
\small
\caption{Fusing the router and the shared expert into the MoE launch. \emph{Fused} (\sysname{}): one launch.
\emph{Separate}: The router (GEMM, top-$k$, aux loss) or the shared expert runs as its own kernels (hipBLASLt for the
shared-expert GEMMs) on the same stream. \emph{Without}: the same MoE layer with routing precomputed (router rows) or
routed experts only (shared-expert rows).}
\label{tab:fusion}
\begin{tabular*}{\textwidth}{@{\extracolsep{\fill}}l l l c c c c@{}}
\toprule
Component & Model & Pass & Fused (ms) & Separate (ms) & Without (ms) & Speedup \\
\midrule
Router        & dsv3    & forward  & \textbf{7.557}  & 7.612  & 7.168  & 1.007 \\
Router        & dsv3    & backward & \textbf{13.773} & 14.005 & 13.711 & 1.017 \\
Router        & qwen235 & forward  & \textbf{3.622}  & 3.679  & 3.398  & 1.016 \\
Router        & qwen235 & backward & \textbf{5.808}  & 5.918  & 5.752  & 1.019 \\
\midrule
Shared expert & dsv3    & forward  & \textbf{7.655}  & 7.990  & 7.218  & 1.044 \\
Shared expert & dsv3    & backward & \textbf{15.533} & 15.797 & 13.865 & 1.017 \\
\bottomrule
\end{tabular*}
\end{table*}
\FloatBarrier
\subsection{Local First AG Queue vs Naive Ordering}
A naive rank-order queue without local-first is 1.17--1.40$\times$ slower than the shipped queue (Table~\ref{tab:agorder}).

\begin{table*}[!ht]
\centering
\small
\caption{AG consumption order, configs 1 (AG+TN) and 3 (AG+NN), TP=8, atomic tile claim (queue order is the
schedule). \emph{Shipped}: local shard first, then peer tiles in the order movers gather them. \emph{Chunk-major}:
local first, peers chunk by chunk (mismatches the gather order). \emph{Naive}: chunk-major in rank order, no
local-first. Time: slowest rank, median of 3 runs $\times$ 25 calls; slowdown vs.\ shipped in parentheses.}
\label{tab:agorder}
\begin{tabular*}{\textwidth}{@{\extracolsep{\fill}}l l c c c@{}}
\toprule
Config & Shape & Shipped (ms) & Chunk-major & Naive \\
\midrule
TN & 70B-FC1-MBS8    & \textbf{7.579}  & 8.040 (+6.1\%)   & 8.873 (+17.1\%)  \\
TN & 405B-QKV-MBS8   & \textbf{5.552}  & 7.314 (+31.7\%)  & 7.742 (+39.4\%)  \\
\midrule
NN & 70B-Down-MBS8   & \textbf{4.038}  & 4.821 (+19.4\%)  & 5.109 (+26.5\%)  \\
NN & 405B-OProj-MBS8 & \textbf{5.184}  & 7.022 (+35.5\%)  & 7.240 (+39.7\%)  \\
\bottomrule
\end{tabular*}
\end{table*}

\FloatBarrier
\subsection{Band-Major vs. Row-Major RS Queue}
Row-major production is 1.41--1.45$\times$ slower than band-major production (Table~\ref{tab:rsorder}).

\begin{table*}[!ht]
\centering
\small
\caption{RS production order, config 2 (GEMM $\rightarrow$ RS), TP=8. Band-major (shipped)
produces band $b$ of every rank's chunk before band $b{+}1$, so reducers start on the first band; row-major
produces each chunk completely before the next.}
\label{tab:rsorder}
\begin{tabular*}{\textwidth}{@{\extracolsep{\fill}}l c c c@{}}
\toprule
Shape & Band-major (ms) & Row-major (ms) & Row / band \\
\midrule
8B-OProj-MBS8  & \textbf{1.366} & 1.930 & 1.41$\times$ \\
70B-Down-MBS8  & \textbf{4.156} & 6.038 & 1.45$\times$ \\
\bottomrule
\end{tabular*}
\end{table*}

\FloatBarrier
\subsection{8 vs. 4 Waves}
With four waves per WG, transport is 1.23--1.96$\times$ slower measured by bytes/CU/s at 7--56 WGs (Table~\ref{tab:waves}).

\begin{table*}[!ht]
\centering
\small
\caption{Mover waves vs.\ request depth, 1\,GiB transfer (70B-MBS8), TP=8. The shipped 8-wave mover keeps one 16-byte
remote load in flight per wave (two waves per single instruction, multiple data unit). 4-wave variants: same loop (depth 1), two loads in flight per wave
(depth 2), or load/store interleave (pipelined, AG only). (Bold: fastest)}
\label{tab:waves}
\begin{tabular*}{\textwidth}{@{\extracolsep{\fill}}l r c c c c@{}}
\toprule
Op & WGs & 8 waves (shipped) & 4 waves & 4 waves, depth 2 & 4 waves, pipelined \\
\midrule
AG & 7   & \textbf{9.772} & 19.131 & 10.525 & 19.460 \\
AG & 14  & \textbf{5.056} & 9.791  & 5.451  & 9.920  \\
AG & 28  & \textbf{2.734} & 5.044  & 2.899  & 5.063  \\
AG & 56  & \textbf{2.217} & 2.727  & 2.220  & 2.729  \\
AG & 112 & 2.278          & \textbf{2.230} & 2.273 & 2.237 \\
\midrule
RS & 7   & \textbf{15.900} & 30.378 & 18.490 & -- \\
RS & 14  & \textbf{7.988}  & 15.205 & 9.263  & -- \\
RS & 28  & \textbf{4.068}  & 7.627  & 4.683  & -- \\
RS & 56  & \textbf{2.288}  & 3.878  & 2.513  & -- \\
RS & 112 & 2.325           & \textbf{2.236} & 2.259 & -- \\
\bottomrule
\end{tabular*}
\end{table*}

\FloatBarrier
\subsection{Dst-Minor vs. Dst-Major Push}
Destination-major push lists are 1.12--2.03$\times$ slower than destination-minor lists (Table~\ref{tab:dstorder}).

\begin{table*}[!ht]
\centering
\small
\caption{Push order for EP transport, dsv3, EP=8, routed experts only (router and shared expert excluded).
\emph{Destination-minor} (\sysname{}): Consecutive WGs push to different destinations, one destination per
WG. \emph{Destination-major}: Consecutive WGs push to the same destination. Same kernels and data.}
\label{tab:dstorder}
\begin{tabular*}{\textwidth}{@{\extracolsep{\fill}}l l c c c@{}}
\toprule
Routing & Phase & Destination-minor (ms) & Destination-major (ms) & Slowdown \\
\midrule
Balanced & Forward  & \textbf{8.148}  & 16.532 & 2.03$\times$ \\
Balanced & Backward & \textbf{13.789} & 20.567 & 1.49$\times$ \\
Skewed   & Forward  & \textbf{18.686} & 24.841 & 1.33$\times$ \\
Skewed   & Backward & \textbf{30.077} & 33.653 & 1.12$\times$ \\
\bottomrule
\end{tabular*}
\end{table*}
\FloatBarrier
\subsection{RS Error vs. RCCL Ring, against a FP64 Reference}
While \sysname{} accumulates all eight partials in FP32 and rounds to BF16 or MXFP8 once, ring algorithms often keep BF16 partials and round at every hop. The result is 2.21$\times$ lower reduction error for \sysname{}, and 1.71-1.73$\times$ lower error against a FP64 GEMM. \sysname{} is also bitwise run-to-run deterministic.

\begin{table*}[!ht]
\centering
\small
\caption{RS error, TP=8. Every arm reduces the same BF16 partials.
\emph{Reduction}: vs.\ the FP64 sum of the partials. \emph{end-to-end}: vs.\ an FP64 GEMM (32{,}768-element sample).}
\label{tab:rserr}
\begin{tabular*}{\textwidth}{@{\extracolsep{\fill}}l ccc ccc@{}}
\toprule
 & \multicolumn{3}{c}{Reduction} & \multicolumn{3}{c}{end-to-end} \\
\cmidrule(lr){2-4}\cmidrule(l){5-7}
Shape & \sysname{} & RCCL & Ratio & \sysname{} & RCCL & Ratio \\
\midrule
8B-OProj-MBS8  & $1.67\times10^{-3}$ & $3.68\times10^{-3}$ & 2.21 & $2.34\times10^{-3}$ & $4.04\times10^{-3}$ & 1.73 \\
70B-Down-MBS8  & $1.67\times10^{-3}$ & $3.68\times10^{-3}$ & 2.21 & $2.35\times10^{-3}$ & $4.02\times10^{-3}$ & 1.71 \\
\bottomrule
\end{tabular*}
\end{table*}

\FloatBarrier
\subsection{Push vs Pull}

While pushing rows into the destination's arena is slightly faster than pulling them this requires more memory than a pull algorithm. As such, for larger EP sizes it may be required to use a pull algorithm instead.

\begin{table*}[!ht]
\centering
\small
\caption{Push vs.\ pull EP transport in the full MoE layer (EP=8). Push (\sysname{}): The producer writes rows into the
destination's buffer. Pull: The consumer reads them from the source in the forward dispatch; combine and the backward push in both arms, so the backward rows are a control. Speedup: pull over push time.}
\label{tab:pushpull}
\begin{tabular*}{\textwidth}{@{\extracolsep{\fill}}l l l c c c@{}}
\toprule
Model & Routing & Phase & Push (ms) & Pull (ms) & Speedup \\
\midrule
dsv3    & balanced & forward  & \textbf{8.117}  & 8.158  & 1.005 \\
dsv3    & balanced & backward & 13.783          & \textbf{13.725} & 0.996 \\
dsv3    & balanced & step     & 21.043          & \textbf{21.025} & 0.999 \\
\midrule
dsv3    & skewed   & forward  & \textbf{18.628} & 19.004 & 1.020 \\
dsv3    & skewed   & backward & 30.055          & \textbf{30.052} & 1.000 \\
dsv3    & skewed   & step     & \textbf{45.144} & 45.506 & 1.008 \\
\midrule
qwen235 & balanced & forward  & \textbf{3.783}  & 3.849  & 1.017 \\
qwen235 & balanced & backward & 5.710           & \textbf{5.699} & 0.998 \\
qwen235 & balanced & step     & \textbf{9.276}  & 9.322  & 1.005 \\
\bottomrule
\end{tabular*}
\end{table*}

\clearpage
\section{Additional \sysname{} Details}
\FloatBarrier
\subsection{Fused Communication Framework Comparison}

\begin{table*}[!ht]
\centering
\scriptsize
\setlength{\tabcolsep}{3pt}
\renewcommand{\arraystretch}{1.1}
\caption{Fused communication--computation systems. \emph{Scope}: what one launch fuses (a collective--GEMM
\emph{pair} or a whole MoE \emph{layer}). \emph{Crew sized by}: how the number of communication units is chosen.
\emph{Roles}/\emph{Switch trigger}: whether and when a unit changes between communication and computation.
\emph{Rtr}/\emph{wgrad}: router and weight gradients inside the launch.}
\label{tab:related}
\setlength{\tabcolsep}{3pt}
\begin{tabularx}{\textwidth}{@{}l >{\raggedright\arraybackslash}p{0.95cm} c c >{\raggedright\arraybackslash}p{2.0cm} >{\raggedright\arraybackslash}p{1.45cm} >{\raggedright\arraybackslash}X c c c c@{}}
\toprule
System & Scope & Train & \shortstack{MoE launches\\F/B} & Crew sized by & Roles & Switch trigger & \shortstack{Switches\\/unit} & Rtr & wgrad & TP+SP \\
\midrule
FLUX~\cite{chang2024fluxfastsoftwarebasedcommunication}               & pair              & \cmark & --          & --                                 & own tile          & --                                                  & 0         & --           & --           & \cmark \\
TileLink~\cite{MLSYS2025_c6ee784c}           & pair              & \xmark & --          & launch config                      & fixed             & --                                                  & 0         & --           & --           & \cmark \\
ParallelKittens~\cite{MLSYS2026_ff997469}    & pair              & \xmark & --          & launch config                      & fixed             & --                                                  & 0         & \xmark       & \xmark       & \cmark \\
Iris~\cite{awad2025irisfirstclassmultigpuprogramming} & pair & \xmark & -- & launch config & fixed & -- & 0 & -- & -- & $\triangle$ \\COMET~\cite{MLSYS2025_e27ea0cd}              & pair              & $\triangle$      & $\sim$2 / ? & offline, per shape                 & fixed             & --                                                  & 0         & \xmark       & ?            & \xmark \\
Primus MegaMoE~\cite{primusturbo_megamoe}     & pair              & \cmark & 3 / 5$^{a}$ & autotuned per shape     & fixed             & --                                                  & 0         & \xmark       & part.$^{a}$  & \xmark \\
UniEP~\cite{10.1145/3806645.3807818}              & pair              & \cmark & 3 / 6$^{a}$ & perf-model tuner                  & one-way           & comm tasks exhausted, queue            & 1         & \xmark       & \xmark$^{a}$ & \xmark \\
FlashMoE~\cite{NEURIPS2025_918d938b}           & layer$^{\dagger}$ & \xmark & 1 / --      & scheduler                          & sched.\ tasks     & scheduler warp                                      & per task  & \cmark       & --           & \xmark \\
DeepSeek MegaMoE~\cite{deepgemm2025}   & layer$^{\dagger}$ & \xmark & 1+aux / --  & launch config                      & fixed warps       & --                                                  & 0         & \xmark       & --           & \xmark \\
Weave~\cite{huang2026weavefinegraineddynamicsm}              & layer$^{\dagger}$ & \xmark & 1+aux / --  & prologue cost model$^{b}$          & planned phases    & precomputed steal count$^{c}$         & $\le$2    & \xmark       & --           & \xmark \\
Mixture-of-Kittens~\cite{sul2026mixtureofkittensmoemegakernelnvl72s} & layer             & \cmark & 1+aux / 1+aux   & launch config           & fixed             & --                                                  & 0         & \xmark$^{d}$ & \cmark       & \xmark \\
MPK~\cite{318585}                & model             & \xmark & 1 / --      & scheduler                          & sched.\ tasks     & scheduler SMs                                       & per task  & \cmark       & --           & \xmark \\
\midrule
\textbf{\sysname{}}    & pair + layer      & \cmark & 1 / 1       & shape model & one-way + two-way & \textbf{readiness at every claim}$^{e}$ & unbounded & \cmark$^{f}$ & \cmark       & \cmark \\
\bottomrule
\end{tabularx}

\vspace{2pt}
\parbox{\textwidth}{\scriptsize
$^{\dagger}$Forward pass only (inference).
$^{a}$From public code.
$^{b}$Per layer and per GPU, from routing counts and calibrated bandwidth/throughput curves; fixed for the rest of the kernel.
$^{c}$Communication SMs dispatch, steal a precomputed number of GEMM tiles, then combine; all SMs join the final combine.
$^{d}$Router runs before the launch; routing-score gradient fused.
$^{e}$Movers hand off when their share is done (AG, dispatch, bulk); combine movers check for ready work before every tile claim, in forward and backward.
$^{f}$Router GEMM, top-$k$ and aux loss forward; router gradients (logits, $dX$, $dW_r$) backward.
}
\end{table*}

\FloatBarrier
\subsection{Supported TP+SP Configs}
\begin{table*}[!ht]
\centering
\caption{TP+SP \sysname{} kernels per collective-GEMM pair ($\dagger$: MXFP8 only).}
\label{tab:tpsp}
\begin{tabular}{llll}
\toprule
Pass & Collective & GEMM & Kernel  \\
\midrule
fwd & AG $X$ & col-parallel & fused AG+TN \\
fwd & RS $Y$ & row-parallel & fused TN+RS \\
bwd & AG $dY$ & row-parallel dgrad & fused AG+NN \\
bwd & AG $dY$, AG $dY^{\top}$\,$\dagger$ & row-parallel dgrad +
 wgrad & fused AG+NN(+NT)\,$\dagger$ \\
bwd & AG $X$ & col-parallel dgrad & bulk AG $\parallel$ NN \\
bwd & RS $dX$ & col-parallel wgrad & bulk RS $\parallel$ NT \\
\bottomrule
\end{tabular}
\end{table*}

\FloatBarrier
\subsection{Heuristics}

\begin{table*}[!ht]
\caption{Per-shape heuristics, evaluated once per shape on the host. Rules were fitted at TP 8 and EP 8; at other TP sizes mover count is scaled. Mover counts are per remote peer; MoE crews are WGs per GPU. All counts are then fitted to the CU cap so that every mover runs beside at least one builder.}
\label{tab:heur}
\vskip 0.1in
\begin{center}
\begin{small}
\begin{tabular}{p{0.9in}p{1.15in}p{1.15in}p{2.85in}}
\toprule
Parameter & Kernel & Input & Rule \\
\midrule
Movers per peer & AG+GEMM, col. fprop & gathered width N & BF16: 6 if N is at least 3584, else 8. MXFP8: 8. \\
 & AG+GEMM, row dgrad & gathered width N & BF16: 6 if N is at least 4608, else 8. MXFP8: 8. \\
 & bulk AG & contraction K & BF16: 2, 4, 6 if K is at least 7168, 3584, 2304; else 8. MXFP8: 2, 4 if K is at least 3584, 1792; else 8. \\
\midrule
Reducers per peer & GEMM+RS, row fprop & local K, waves w (tiles per cap) & K at least 4096: 4 (MXFP8 5). K at least 3072: 6 (MXFP8 8 if w is at least 16, else 7). K at most 768: 14, 10, 8 if w is at least 16, at least 8, or less. Otherwise 8 (MXFP8 10 if w is at least 16). \\
 & bulk RS & output width n & BF16: 2, 4 if n is at least 8192, 3584; else 8. MXFP8: 8 if n is 5120 to 8191, else 4. \\
\midrule
Dispatch crew & MoE forward & expert width I & BF16: 32, 24, 16 if I is below 1024, below 3072, or larger. MXFP8: 24576/I rounded to a multiple of 8 between 16 and 48, doubled if the forward also sends the column-wise input. Minimum over ranks. \\
Combine crew & MoE forward & CU cap & The smaller of 128 and the cap, rounded down to a multiple of 8. \\
Dispatch crew & MoE backward & total bytes, total GEMM work & 256 WGs times push time over push plus GEMM time, rounded up to a multiple of 8, between 32 and 64. \\
Combine crew & MoE backward & total bytes, total GEMM work & 256 WGs times push time over GEMM time, rounded up to a multiple of 8, between 24 and 128 (MXFP8 32 and 128). \\
\midrule
Push burst depth & MoE dispatch, combine & hidden size H & Largest of 8, 7, 6, 4 that divides H/512. \\
\midrule
Split-K factor & AG+GEMM row dgrad, bulk AG & output tiles & BF16: 4 or 2 if the FP32 partials fit the workspace (at most 64 MiB), else 1. MXFP8: 1. \\
 & bulk RS & output tiles & BF16: 4 if at most 64 tiles and the partials fit the workspace, else 1. MXFP8: 1. \\
\midrule
XCD bucketing & AG+GEMM (all) & tiles, tile rows & On if at most 1024 tiles and at least 5 tile rows. \\
\bottomrule
\end{tabular}
\end{small}
\end{center}
\vskip -0.1in
\end{table*}

\clearpage
\section{Full Per-Shape and Per-GPU Tables behind the Summary Figures}

\begin{table*}[!ht]
\caption{Llama~3 TP+SP end-to-end training, TP~8, sequence 8192, GBS~64: median over three runs of Megatron-LM's printed numbers (each the mean over iterations 3--19). Speedup is TE baseline ms/iter over \sysname{} ms/iter. OOM: out of memory. \textsuperscript{r}Full recompute in both arms. Layer-reduced proxies: The 405B model runs 24 of its 126 layers with full recompute. 70B at MBS~4 and~8 runs with full recompute. Three runs per point, 20 iterations each.}
\label{tab:e2e-tpsp}
\vskip 0.1in
\begin{center}
\begin{footnotesize}
\setlength{\tabcolsep}{4pt}
\begin{tabular}{lrrrrrrrrrr}
\toprule
 & & & \multicolumn{4}{c}{BF16} & \multicolumn{4}{c}{MXFP8} \\
\cmidrule(lr){4-7}\cmidrule(lr){8-11}
Model & MBS & Layers & TE baseline & \sysname{} & \sysname{} & Speedup & TE baseline & \sysname{} & \sysname{} & Speedup \\
 & & & ms/iter & ms/iter & TFLOPS & & ms/iter & ms/iter & TFLOPS & \\
\midrule
\multicolumn{11}{l}{\textbf{MI350X}} \\
8B & 1 & 32 & 10056.2 & 7913.1 & 426.6 & 1.271 & 8708.3 & 6967.7 & 484.9 & 1.250 \\
8B & 2 & 32 & 7727.6 & 6435.7 & 524.4 & 1.201 & 6729.2 & 5454.8 & 619.1 & 1.234 \\
8B & 4 & 32 & 7119.4 & 5811.7 & 580.7 & 1.225 & 5939.6 & 4902.3 & 688.5 & 1.212 \\
8B & 8 & 32 & 7117.2 & 5658.0 & 596.4 & 1.258 & 5813.0 & 4702.4 & 717.6 & 1.236 \\
70B & 1 & 80 & 47371.0 & 41279.4 & 713.3 & 1.148 & 37514.8 & 30841.8 & 954.7 & 1.216 \\
70B & 2 & 80 & 44371.2 & 38308.3 & 768.6 & 1.158 & 32956.2 & 27603.6 & 1066.6 & 1.194 \\
70B & 4 & 80\textsuperscript{r} & 62232.3 & 51524.3 & 571.4 & 1.208 & 43348.7 & 36146.8 & 814.5 & 1.199 \\
70B & 8 & 80\textsuperscript{r} & 62377.7 & 51415.7 & 572.6 & 1.213 & 43643.7 & 36258.4 & 812.0 & 1.204 \\
405B & 1 & 24\textsuperscript{r} & 56475.7 & 51236.1 & 628.1 & 1.102 & 37706.9 & 34361.3 & 936.5 & 1.097 \\
405B & 2 & 24\textsuperscript{r} & 57459.5 & 49923.4 & 644.6 & 1.151 & 37438.2 & 32840.9 & 979.9 & 1.140 \\
405B & 4 & 24\textsuperscript{r} & 57912.5 & 49899.2 & 644.9 & 1.161 & 37347.7 & 32671.8 & 985.0 & 1.143 \\
405B & 8 & 24\textsuperscript{r} & 58049.1 & 49998.7 & 643.6 & 1.161 & 37655.0 & OOM & -- & -- \\
\midrule
\multicolumn{11}{l}{\textbf{MI355X}} \\
8B & 1 & 32 & 9996.1 & 7915.3 & 427.0 & 1.263 & 8647.1 & 6971.3 & 485.5 & 1.240 \\
8B & 2 & 32 & 7876.5 & 6351.8 & 531.3 & 1.240 & 6664.8 & 5368.7 & 628.9 & 1.241 \\
8B & 4 & 32 & 7220.7 & 5709.9 & 591.0 & 1.265 & 5893.8 & 4833.4 & 698.5 & 1.219 \\
8B & 8 & 32 & 7059.0 & 5515.6 & 611.8 & 1.280 & 5649.1 & 4619.6 & 730.5 & 1.223 \\
70B & 1 & 80 & 48695.2 & 39918.9 & 737.6 & 1.220 & 37191.5 & 30146.4 & 976.7 & 1.234 \\
70B & 2 & 80 & 45029.7 & 35307.9 & 833.9 & 1.275 & 32447.1 & 26317.1 & 1118.8 & 1.233 \\
70B & 4 & 80\textsuperscript{r} & 59942.3 & 47738.9 & 616.7 & 1.256 & 42303.4 & 34680.9 & 848.9 & 1.220 \\
70B & 8 & 80\textsuperscript{r} & 58645.8 & 46949.1 & 627.1 & 1.249 & 41509.8 & 34187.4 & 861.2 & 1.214 \\
405B & 1 & 24\textsuperscript{r} & 54176.8 & 46203.8 & 696.5 & 1.173 & 37137.2 & 31882.3 & 1009.4 & 1.165 \\
405B & 2 & 24\textsuperscript{r} & 52786.9 & 43727.3 & 735.9 & 1.207 & 35251.2 & 29995.2 & 1072.9 & 1.175 \\
405B & 4 & 24\textsuperscript{r} & 52032.3 & 42743.1 & 752.9 & 1.217 & 34474.2 & 29407.8 & 1094.3 & 1.172 \\
405B & 8 & 24\textsuperscript{r} & 51565.3 & 42513.7 & 756.9 & 1.213 & 33989.8 & 29226.6 & 1101.1 & 1.163 \\
\bottomrule
\end{tabular}
\end{footnotesize}
\end{center}
\vskip -0.1in
\end{table*}

\begin{table*}[p]
\caption{MoE proxy training, EP~8: median over three runs of Megatron-LM's printed numbers, 20 iterations each. \sysname{} speedup is the DeepEP ms/iter over \sysname{}'s. DeepEP: Megatron-LM with ROCm's DeepEP. Balanced: uniform random routing; skewed: the model's top-k router.}
\label{tab:e2e-moe}
\vskip 0.1in
\begin{center}
\begin{footnotesize}
\setlength{\tabcolsep}{4pt}
\begin{tabular}{lllrrrrrr}
\toprule
 & & & \multicolumn{3}{c}{BF16} & \multicolumn{3}{c}{MXFP8} \\
\cmidrule(lr){4-6}\cmidrule(lr){7-9}
Model & Routing & Backend & ms/iter & TFLOPS & \sysname{} speedup & ms/iter & TFLOPS & \sysname{} speedup \\
\midrule
\multicolumn{9}{l}{\textbf{MI350X}} \\
Qwen3-235B & balanced & \textbf{\sysname{}} & 545.1 & 610 & -- & 491.4 & 677 & -- \\
 & & DeepEP & 925.6 & 359 & 1.698 & 894.7 & 372 & 1.821 \\
Qwen3-235B & skewed & \textbf{\sysname{}} & 1022.6 & 327 & -- & 868.1 & 384 & -- \\
 & & DeepEP & 1337.7 & 249 & 1.308 & 1359.2 & 245 & 1.566 \\
DeepSeek-V3 & balanced & \textbf{\sysname{}} & 612.7 & 556 & -- & 550.1 & 619 & -- \\
 & & DeepEP & 804.8 & 423 & 1.314 & 762.8 & 446 & 1.387 \\
DeepSeek-V3 & skewed & \textbf{\sysname{}} & 617.1 & 551 & -- & 555.4 & 613 & -- \\
 & & DeepEP & 822.7 & 414 & 1.333 & 778.5 & 437 & 1.402 \\
\midrule
\multicolumn{9}{l}{\textbf{MI355X}} \\
Qwen3-235B & balanced & \textbf{\sysname{}} & 494.3 & 673 & -- & 454.9 & 731 & -- \\
 & & DeepEP & 859.7 & 387 & 1.739 & 831.1 & 400 & 1.827 \\
Qwen3-235B & skewed & \textbf{\sysname{}} & 941.6 & 355 & -- & 810.5 & 411 & -- \\
 & & DeepEP & 1211.8 & 275 & 1.287 & 1244.2 & 268 & 1.535 \\
DeepSeek-V3 & balanced & \textbf{\sysname{}} & 534.3 & 637 & -- & 488.7 & 696 & -- \\
 & & DeepEP & 709.7 & 480 & 1.328 & 680.0 & 500 & 1.391 \\
DeepSeek-V3 & skewed & \textbf{\sysname{}} & 539.7 & 631 & -- & 493.8 & 689 & -- \\
 & & DeepEP & 732.1 & 465 & 1.356 & 697.9 & 488 & 1.413 \\
\bottomrule
\end{tabular}
\end{footnotesize}
\end{center}
\vskip -0.1in
\end{table*}

\begin{table*}[p]
\caption{TP+SP microbenchmark, AG+GEMM, column fprop, TP~8, MI350X: per iteration the slowest rank, median. Base: the RCCL collective and the GEMM run back to back on one stream with no overlap (BF16: torch.mm (hipBLASLt); MXFP8: a communication-free HipKittens MXFP8 GEMM). Speedup is base over \sysname{}; vs unf.\ is \sysname{}'s communication and GEMM WGs run back to back over the fused kernel.}
\label{tab:micro-tpsp-cfg1-MI350X}
\vskip 0.1in
\begin{center}
\begin{footnotesize}
\setlength{\tabcolsep}{3pt}
\begin{tabular}{lrrrrrrrrrrr}
\toprule
 & & & & \multicolumn{4}{c}{BF16} & \multicolumn{4}{c}{MXFP8} \\
\cmidrule(lr){5-8}\cmidrule(lr){9-12}
Shape & M & N & K & Base ms & \sysname{} ms & Speedup & vs unf. & Base ms & \sysname{} ms & Speedup & vs unf. \\
\midrule
405B-FC1-MBS1 & 8192 & 13312 & 16384 & 3.994 & 3.320 & 1.203 & 1.188 & 1.903 & 1.651 & 1.152 & 1.100 \\
405B-QKV-MBS1 & 8192 & 2304 & 16384 & 1.128 & 0.926 & 1.218 & 1.366 & 0.740 & 0.521 & 1.420 & 1.343 \\
70B-FC1-MBS1 & 8192 & 7168 & 8192 & 1.091 & 0.995 & 1.097 & 1.072 & 0.613 & 0.539 & 1.137 & 1.134 \\
70B-QKV-MBS1 & 8192 & 1280 & 8192 & 0.516 & 0.469 & 1.101 & 1.028 & 0.308 & 0.281 & 1.097 & 1.044 \\
8B-FC1-MBS1 & 8192 & 3584 & 4096 & 0.502 & 0.310 & 1.619 & 1.241 & 0.259 & 0.209 & 1.242 & 1.180 \\
8B-QKV-MBS1 & 8192 & 768 & 4096 & 0.377 & 0.234 & 1.611 & 1.041 & 0.192 & 0.165 & 1.167 & 1.040 \\
405B-FC1-MBS2 & 16384 & 13312 & 16384 & 8.108 & 6.908 & 1.174 & 1.137 & 3.943 & 3.476 & 1.134 & 1.086 \\
405B-QKV-MBS2 & 16384 & 2304 & 16384 & 2.170 & 1.569 & 1.383 & 1.434 & 1.349 & 0.883 & 1.529 & 1.413 \\
70B-FC1-MBS2 & 16384 & 7168 & 8192 & 1.963 & 1.734 & 1.132 & 1.165 & 1.092 & 0.966 & 1.132 & 1.181 \\
70B-QKV-MBS2 & 16384 & 1280 & 8192 & 0.993 & 0.756 & 1.313 & 1.262 & 0.583 & 0.442 & 1.318 & 1.240 \\
8B-FC1-MBS2 & 16384 & 3584 & 4096 & 0.766 & 0.544 & 1.409 & 1.334 & 0.444 & 0.346 & 1.282 & 1.265 \\
8B-QKV-MBS2 & 16384 & 768 & 4096 & 0.468 & 0.398 & 1.174 & 1.036 & 0.269 & 0.248 & 1.083 & 1.048 \\
405B-FC1-MBS4 & 32768 & 13312 & 16384 & 16.169 & 13.525 & 1.196 & 1.123 & 8.418 & 7.455 & 1.129 & 1.067 \\
405B-QKV-MBS4 & 32768 & 2304 & 16384 & 4.268 & 2.832 & 1.507 & 1.515 & 2.493 & 1.589 & 1.568 & 1.463 \\
70B-FC1-MBS4 & 32768 & 7168 & 8192 & 4.664 & 3.530 & 1.321 & 1.237 & 2.174 & 1.894 & 1.148 & 1.190 \\
70B-QKV-MBS4 & 32768 & 1280 & 8192 & 1.832 & 1.353 & 1.354 & 1.288 & 1.066 & 0.786 & 1.357 & 1.277 \\
8B-FC1-MBS4 & 32768 & 3584 & 4096 & 1.348 & 0.938 & 1.436 & 1.460 & 0.822 & 0.619 & 1.328 & 1.295 \\
8B-QKV-MBS4 & 32768 & 768 & 4096 & 0.873 & 0.693 & 1.260 & 1.194 & 0.489 & 0.411 & 1.189 & 1.177 \\
405B-FC1-MBS8 & 65536 & 13312 & 16384 & 31.274 & 26.651 & 1.173 & 1.131 & 16.921 & 14.983 & 1.129 & 1.085 \\
405B-QKV-MBS8 & 65536 & 2304 & 16384 & 9.082 & 5.578 & 1.628 & 1.612 & 4.993 & 3.116 & 1.603 & 1.485 \\
70B-FC1-MBS8 & 65536 & 7168 & 8192 & 10.035 & 7.678 & 1.307 & 1.257 & 5.006 & 4.238 & 1.181 & 1.159 \\
70B-QKV-MBS8 & 65536 & 1280 & 8192 & 3.528 & 2.532 & 1.394 & 1.344 & 2.074 & 1.479 & 1.402 & 1.317 \\
8B-FC1-MBS8 & 65536 & 3584 & 4096 & 2.751 & 1.829 & 1.505 & 1.536 & 1.621 & 1.225 & 1.324 & 1.280 \\
8B-QKV-MBS8 & 65536 & 768 & 4096 & 1.668 & 1.289 & 1.294 & 1.219 & 0.919 & 0.766 & 1.200 & 1.199 \\
\emph{geometric mean} & & & & & & \emph{1.316} & \emph{1.249} & & & \emph{1.252} & \emph{1.205} \\
\bottomrule
\end{tabular}
\end{footnotesize}
\end{center}
\vskip -0.1in
\end{table*}

\begin{table*}[p]
\caption{TP+SP microbenchmark, AG+GEMM, column fprop, TP~8, MI355X: per iteration the slowest rank, median. Base: the RCCL collective and the GEMM run back to back on one stream with no overlap (BF16: torch.mm (hipBLASLt); MXFP8: a communication-free HipKittens MXFP8 GEMM). Speedup is base over \sysname{}; vs unf.\ is \sysname{}'s communication and GEMM WGs run back to back over the fused kernel.}
\label{tab:micro-tpsp-cfg1-MI355X}
\vskip 0.1in
\begin{center}
\begin{footnotesize}
\setlength{\tabcolsep}{3pt}
\begin{tabular}{lrrrrrrrrrrr}
\toprule
 & & & & \multicolumn{4}{c}{BF16} & \multicolumn{4}{c}{MXFP8} \\
\cmidrule(lr){5-8}\cmidrule(lr){9-12}
Shape & M & N & K & Base ms & \sysname{} ms & Speedup & vs unf. & Base ms & \sysname{} ms & Speedup & vs unf. \\
\midrule
405B-FC1-MBS1 & 8192 & 13312 & 16384 & 3.705 & 2.929 & 1.265 & 1.239 & 1.958 & 1.625 & 1.205 & 1.202 \\
405B-QKV-MBS1 & 8192 & 2304 & 16384 & 1.200 & 0.959 & 1.252 & 1.431 & 0.754 & 0.524 & 1.438 & 1.419 \\
70B-FC1-MBS1 & 8192 & 7168 & 8192 & 1.183 & 0.958 & 1.235 & 1.261 & 0.656 & 0.532 & 1.233 & 1.251 \\
70B-QKV-MBS1 & 8192 & 1280 & 8192 & 0.540 & 0.486 & 1.113 & 1.031 & 0.308 & 0.291 & 1.059 & 1.051 \\
8B-FC1-MBS1 & 8192 & 3584 & 4096 & 0.499 & 0.308 & 1.621 & 1.328 & 0.258 & 0.210 & 1.229 & 1.192 \\
8B-QKV-MBS1 & 8192 & 768 & 4096 & 0.337 & 0.243 & 1.386 & 1.042 & 0.190 & 0.169 & 1.125 & 1.030 \\
405B-FC1-MBS2 & 16384 & 13312 & 16384 & 7.427 & 5.848 & 1.270 & 1.208 & 3.735 & 3.176 & 1.176 & 1.159 \\
405B-QKV-MBS2 & 16384 & 2304 & 16384 & 2.335 & 1.639 & 1.424 & 1.544 & 1.340 & 0.910 & 1.472 & 1.491 \\
70B-FC1-MBS2 & 16384 & 7168 & 8192 & 2.289 & 1.682 & 1.361 & 1.345 & 1.226 & 1.000 & 1.227 & 1.247 \\
70B-QKV-MBS2 & 16384 & 1280 & 8192 & 1.008 & 0.800 & 1.260 & 1.318 & 0.580 & 0.455 & 1.276 & 1.270 \\
8B-FC1-MBS2 & 16384 & 3584 & 4096 & 0.795 & 0.536 & 1.482 & 1.481 & 0.461 & 0.347 & 1.328 & 1.319 \\
8B-QKV-MBS2 & 16384 & 768 & 4096 & 0.483 & 0.421 & 1.149 & 1.039 & 0.271 & 0.257 & 1.053 & 1.045 \\
405B-FC1-MBS4 & 32768 & 13312 & 16384 & 14.025 & 11.265 & 1.245 & 1.229 & 7.511 & 6.332 & 1.186 & 1.154 \\
405B-QKV-MBS4 & 32768 & 2304 & 16384 & 4.648 & 2.924 & 1.590 & 1.581 & 2.525 & 1.664 & 1.518 & 1.538 \\
70B-FC1-MBS4 & 32768 & 7168 & 8192 & 4.490 & 3.206 & 1.401 & 1.371 & 2.393 & 1.923 & 1.245 & 1.271 \\
70B-QKV-MBS4 & 32768 & 1280 & 8192 & 1.926 & 1.433 & 1.344 & 1.347 & 1.050 & 0.816 & 1.288 & 1.316 \\
8B-FC1-MBS4 & 32768 & 3584 & 4096 & 1.534 & 0.943 & 1.626 & 1.624 & 0.847 & 0.620 & 1.366 & 1.360 \\
8B-QKV-MBS4 & 32768 & 768 & 4096 & 0.891 & 0.734 & 1.215 & 1.186 & 0.497 & 0.430 & 1.155 & 1.172 \\
405B-FC1-MBS8 & 65536 & 13312 & 16384 & 27.347 & 21.930 & 1.247 & 1.234 & 14.561 & 12.447 & 1.170 & 1.157 \\
405B-QKV-MBS8 & 65536 & 2304 & 16384 & 9.179 & 5.396 & 1.701 & 1.668 & 4.892 & 3.205 & 1.526 & 1.545 \\
70B-FC1-MBS8 & 65536 & 7168 & 8192 & 9.132 & 6.475 & 1.410 & 1.347 & 4.776 & 3.738 & 1.278 & 1.266 \\
70B-QKV-MBS8 & 65536 & 1280 & 8192 & 3.703 & 2.686 & 1.378 & 1.359 & 2.036 & 1.551 & 1.313 & 1.319 \\
8B-FC1-MBS8 & 65536 & 3584 & 4096 & 2.968 & 1.808 & 1.641 & 1.624 & 1.661 & 1.250 & 1.329 & 1.349 \\
8B-QKV-MBS8 & 65536 & 768 & 4096 & 1.720 & 1.375 & 1.251 & 1.205 & 0.926 & 0.802 & 1.155 & 1.182 \\
\emph{geometric mean} & & & & & & \emph{1.360} & \emph{1.323} & & & \emph{1.258} & \emph{1.255} \\
\bottomrule
\end{tabular}
\end{footnotesize}
\end{center}
\vskip -0.1in
\end{table*}

\clearpage

\begin{table*}[p]
\caption{TP+SP microbenchmark, GEMM+RS, row fprop, TP~8, MI350X: per iteration the slowest rank, median. Base: the RCCL collective and the GEMM run back to back on one stream with no overlap (BF16: torch.mm (hipBLASLt); MXFP8: a communication-free HipKittens MXFP8 GEMM). Speedup is base over \sysname{}; vs unf.\ is \sysname{}'s communication and GEMM WGs run back to back over the fused kernel.}
\label{tab:micro-tpsp-cfg2-MI350X}
\vskip 0.1in
\begin{center}
\begin{footnotesize}
\setlength{\tabcolsep}{3pt}
\begin{tabular}{lrrrrrrrrrrr}
\toprule
 & & & & \multicolumn{4}{c}{BF16} & \multicolumn{4}{c}{MXFP8} \\
\cmidrule(lr){5-8}\cmidrule(lr){9-12}
Shape & M & N & K & Base ms & \sysname{} ms & Speedup & vs unf. & Base ms & \sysname{} ms & Speedup & vs unf. \\
\midrule
405B-Down-MBS1 & 8192 & 16384 & 6656 & 2.015 & 1.703 & 1.183 & 1.252 & 1.382 & 1.253 & 1.103 & 1.396 \\
405B-OProj-MBS1 & 8192 & 16384 & 2048 & 1.140 & 0.819 & 1.392 & 1.460 & 0.967 & 0.766 & 1.261 & 1.367 \\
70B-Down-MBS1 & 8192 & 8192 & 3584 & 0.752 & 0.528 & 1.424 & 1.316 & 0.576 & 0.441 & 1.307 & 1.306 \\
70B-OProj-MBS1 & 8192 & 8192 & 1024 & 0.541 & 0.375 & 1.441 & 1.297 & 0.462 & 0.381 & 1.215 & 1.220 \\
8B-Down-MBS1 & 8192 & 4096 & 1792 & 0.354 & 0.234 & 1.509 & 1.317 & 0.271 & 0.222 & 1.217 & 1.238 \\
8B-OProj-MBS1 & 8192 & 4096 & 512 & 0.293 & 0.205 & 1.431 & 1.246 & 0.239 & 0.210 & 1.138 & 1.203 \\
405B-Down-MBS2 & 16384 & 16384 & 6656 & 4.411 & 3.646 & 1.210 & 1.291 & 2.728 & 2.401 & 1.136 & 1.464 \\
405B-OProj-MBS2 & 16384 & 16384 & 2048 & 2.252 & 1.581 & 1.424 & 1.487 & 1.882 & 1.417 & 1.328 & 1.381 \\
70B-Down-MBS2 & 16384 & 8192 & 3584 & 1.389 & 1.002 & 1.386 & 1.453 & 1.105 & 0.792 & 1.395 & 1.560 \\
70B-OProj-MBS2 & 16384 & 8192 & 1024 & 0.971 & 0.712 & 1.364 & 1.371 & 0.877 & 0.718 & 1.221 & 1.333 \\
8B-Down-MBS2 & 16384 & 4096 & 1792 & 0.593 & 0.381 & 1.555 & 1.477 & 0.502 & 0.384 & 1.309 & 1.327 \\
8B-OProj-MBS2 & 16384 & 4096 & 512 & 0.503 & 0.356 & 1.413 & 1.309 & 0.440 & 0.379 & 1.161 & 1.240 \\
405B-Down-MBS4 & 32768 & 16384 & 6656 & 9.581 & 7.571 & 1.265 & 1.287 & 5.713 & 4.763 & 1.199 & 1.538 \\
405B-OProj-MBS4 & 32768 & 16384 & 2048 & 4.385 & 3.087 & 1.421 & 1.528 & 3.674 & 2.723 & 1.349 & 1.398 \\
70B-Down-MBS4 & 32768 & 8192 & 3584 & 2.837 & 1.880 & 1.509 & 1.601 & 2.167 & 1.436 & 1.509 & 1.650 \\
70B-OProj-MBS4 & 32768 & 8192 & 1024 & 1.886 & 1.446 & 1.304 & 1.298 & 1.700 & 1.350 & 1.259 & 1.280 \\
8B-Down-MBS4 & 32768 & 4096 & 1792 & 1.118 & 0.714 & 1.567 & 1.607 & 0.960 & 0.711 & 1.350 & 1.469 \\
8B-OProj-MBS4 & 32768 & 4096 & 512 & 0.864 & 0.698 & 1.238 & 1.307 & 0.837 & 0.721 & 1.161 & 1.257 \\
405B-Down-MBS8 & 65536 & 16384 & 6656 & 18.697 & 15.043 & 1.243 & 1.284 & 11.873 & 9.270 & 1.281 & 1.597 \\
405B-OProj-MBS8 & 65536 & 16384 & 2048 & 9.350 & 6.196 & 1.509 & 1.602 & 7.322 & 5.297 & 1.382 & 1.434 \\
70B-Down-MBS8 & 65536 & 8192 & 3584 & 6.004 & 4.158 & 1.444 & 1.527 & 4.271 & 2.894 & 1.476 & 1.566 \\
70B-OProj-MBS8 & 65536 & 8192 & 1024 & 3.621 & 2.921 & 1.240 & 1.260 & 3.313 & 2.629 & 1.260 & 1.278 \\
8B-Down-MBS8 & 65536 & 4096 & 1792 & 2.171 & 1.444 & 1.504 & 1.548 & 1.863 & 1.341 & 1.390 & 1.434 \\
8B-OProj-MBS8 & 65536 & 4096 & 512 & 1.704 & 1.367 & 1.247 & 1.350 & 1.622 & 1.361 & 1.191 & 1.306 \\
\emph{geometric mean} & & & & & & \emph{1.380} & \emph{1.390} & & & \emph{1.271} & \emph{1.379} \\
\bottomrule
\end{tabular}
\end{footnotesize}
\end{center}
\vskip -0.1in
\end{table*}

\begin{table*}[p]
\caption{TP+SP microbenchmark, GEMM+RS, row fprop, TP~8, MI355X: per iteration the slowest rank, median. Base: the RCCL collective and the GEMM run back to back on one stream with no overlap (BF16: torch.mm (hipBLASLt); MXFP8: a communication-free HipKittens MXFP8 GEMM). Speedup is base over \sysname{}; vs unf.\ is \sysname{}'s communication and GEMM WGs run back to back over the fused kernel.}
\label{tab:micro-tpsp-cfg2-MI355X}
\vskip 0.1in
\begin{center}
\begin{footnotesize}
\setlength{\tabcolsep}{3pt}
\begin{tabular}{lrrrrrrrrrrr}
\toprule
 & & & & \multicolumn{4}{c}{BF16} & \multicolumn{4}{c}{MXFP8} \\
\cmidrule(lr){5-8}\cmidrule(lr){9-12}
Shape & M & N & K & Base ms & \sysname{} ms & Speedup & vs unf. & Base ms & \sysname{} ms & Speedup & vs unf. \\
\midrule
405B-Down-MBS1 & 8192 & 16384 & 6656 & 2.164 & 1.795 & 1.206 & 1.271 & 1.514 & 1.184 & 1.279 & 1.525 \\
405B-OProj-MBS1 & 8192 & 16384 & 2048 & 1.194 & 0.840 & 1.421 & 1.433 & 1.016 & 0.770 & 1.320 & 1.338 \\
70B-Down-MBS1 & 8192 & 8192 & 3584 & 0.807 & 0.559 & 1.445 & 1.378 & 0.609 & 0.442 & 1.377 & 1.383 \\
70B-OProj-MBS1 & 8192 & 8192 & 1024 & 0.539 & 0.393 & 1.371 & 1.307 & 0.476 & 0.383 & 1.242 & 1.212 \\
8B-Down-MBS1 & 8192 & 4096 & 1792 & 0.353 & 0.246 & 1.432 & 1.308 & 0.275 & 0.229 & 1.200 & 1.218 \\
8B-OProj-MBS1 & 8192 & 4096 & 512 & 0.281 & 0.219 & 1.283 & 1.220 & 0.244 & 0.220 & 1.107 & 1.173 \\
405B-Down-MBS2 & 16384 & 16384 & 6656 & 4.309 & 3.320 & 1.298 & 1.344 & 2.973 & 2.198 & 1.352 & 1.603 \\
405B-OProj-MBS2 & 16384 & 16384 & 2048 & 2.404 & 1.530 & 1.571 & 1.545 & 1.986 & 1.454 & 1.366 & 1.374 \\
70B-Down-MBS2 & 16384 & 8192 & 3584 & 1.520 & 1.035 & 1.468 & 1.506 & 1.181 & 0.792 & 1.491 & 1.572 \\
70B-OProj-MBS2 & 16384 & 8192 & 1024 & 0.995 & 0.735 & 1.354 & 1.329 & 0.914 & 0.721 & 1.268 & 1.286 \\
8B-Down-MBS2 & 16384 & 4096 & 1792 & 0.620 & 0.401 & 1.545 & 1.493 & 0.517 & 0.388 & 1.334 & 1.324 \\
8B-OProj-MBS2 & 16384 & 4096 & 512 & 0.483 & 0.383 & 1.261 & 1.267 & 0.452 & 0.384 & 1.176 & 1.225 \\
405B-Down-MBS4 & 32768 & 16384 & 6656 & 8.764 & 6.467 & 1.355 & 1.351 & 5.808 & 4.365 & 1.331 & 1.596 \\
405B-OProj-MBS4 & 32768 & 16384 & 2048 & 4.688 & 2.949 & 1.590 & 1.549 & 3.893 & 2.776 & 1.402 & 1.400 \\
70B-Down-MBS4 & 32768 & 8192 & 3584 & 3.047 & 1.911 & 1.594 & 1.630 & 2.315 & 1.458 & 1.588 & 1.588 \\
70B-OProj-MBS4 & 32768 & 8192 & 1024 & 1.972 & 1.456 & 1.354 & 1.308 & 1.777 & 1.393 & 1.276 & 1.274 \\
8B-Down-MBS4 & 32768 & 4096 & 1792 & 1.154 & 0.740 & 1.559 & 1.579 & 1.002 & 0.722 & 1.388 & 1.417 \\
8B-OProj-MBS4 & 32768 & 4096 & 512 & 0.892 & 0.735 & 1.214 & 1.295 & 0.870 & 0.744 & 1.169 & 1.234 \\
405B-Down-MBS8 & 65536 & 16384 & 6656 & 16.904 & 12.565 & 1.345 & 1.380 & 11.491 & 8.499 & 1.352 & 1.615 \\
405B-OProj-MBS8 & 65536 & 16384 & 2048 & 9.329 & 5.796 & 1.610 & 1.612 & 7.697 & 5.408 & 1.423 & 1.433 \\
70B-Down-MBS8 & 65536 & 8192 & 3584 & 5.955 & 3.731 & 1.596 & 1.613 & 4.547 & 2.861 & 1.589 & 1.595 \\
70B-OProj-MBS8 & 65536 & 8192 & 1024 & 3.815 & 2.906 & 1.313 & 1.278 & 3.472 & 2.719 & 1.277 & 1.285 \\
8B-Down-MBS8 & 65536 & 4096 & 1792 & 2.280 & 1.495 & 1.525 & 1.510 & 1.959 & 1.379 & 1.420 & 1.425 \\
8B-OProj-MBS8 & 65536 & 4096 & 512 & 1.768 & 1.440 & 1.228 & 1.335 & 1.695 & 1.433 & 1.183 & 1.294 \\
\emph{geometric mean} & & & & & & \emph{1.408} & \emph{1.404} & & & \emph{1.324} & \emph{1.384} \\
\bottomrule
\end{tabular}
\end{footnotesize}
\end{center}
\vskip -0.1in
\end{table*}

\begin{table*}[p]
\caption{TP+SP microbenchmark, AG+GEMM, row dgrad, TP~8, MI350X: per iteration the slowest rank, median. Base: the RCCL collective and the GEMM run back to back on one stream with no overlap (BF16: torch.mm (hipBLASLt); MXFP8: a communication-free HipKittens MXFP8 GEMM). Speedup is base over \sysname{}; vs unf.\ is \sysname{}'s communication and GEMM WGs run back to back over the fused kernel.}
\label{tab:micro-tpsp-cfg3-MI350X}
\vskip 0.1in
\begin{center}
\begin{footnotesize}
\setlength{\tabcolsep}{3pt}
\begin{tabular}{lrrrrrrrrrrr}
\toprule
 & & & & \multicolumn{4}{c}{BF16} & \multicolumn{4}{c}{MXFP8} \\
\cmidrule(lr){5-8}\cmidrule(lr){9-12}
Shape & M & N & K & Base ms & \sysname{} ms & Speedup & vs unf. & Base ms & \sysname{} ms & Speedup & vs unf. \\
\midrule
405B-Down-MBS1 & 8192 & 6656 & 16384 & 2.172 & 1.976 & 1.099 & 1.127 & 2.314 & 1.840 & 1.258 & 1.163 \\
405B-OProj-MBS1 & 8192 & 2048 & 16384 & 1.069 & 0.940 & 1.137 & 1.070 & 1.171 & 0.871 & 1.345 & 1.269 \\
70B-Down-MBS1 & 8192 & 3584 & 8192 & 0.743 & 0.666 & 1.115 & 1.074 & 0.806 & 0.582 & 1.384 & 1.377 \\
70B-OProj-MBS1 & 8192 & 1024 & 8192 & 0.511 & 0.421 & 1.214 & 1.056 & 0.660 & 0.382 & 1.728 & 1.240 \\
8B-Down-MBS1 & 8192 & 1792 & 4096 & 0.411 & 0.273 & 1.506 & 1.025 & 0.465 & 0.269 & 1.730 & 1.386 \\
8B-OProj-MBS1 & 8192 & 512 & 4096 & 0.368 & 0.220 & 1.674 & 1.152 & 0.312 & 0.238 & 1.307 & 1.101 \\
405B-Down-MBS2 & 16384 & 6656 & 16384 & 4.209 & 3.696 & 1.139 & 1.215 & 4.916 & 4.165 & 1.180 & 1.145 \\
405B-OProj-MBS2 & 16384 & 2048 & 16384 & 2.052 & 1.529 & 1.343 & 1.299 & 2.267 & 1.742 & 1.301 & 1.264 \\
70B-Down-MBS2 & 16384 & 3584 & 8192 & 1.430 & 1.160 & 1.233 & 1.195 & 1.531 & 1.088 & 1.407 & 1.397 \\
70B-OProj-MBS2 & 16384 & 1024 & 8192 & 0.897 & 0.764 & 1.175 & 1.055 & 1.074 & 0.680 & 1.579 & 1.442 \\
8B-Down-MBS2 & 16384 & 1792 & 4096 & 0.575 & 0.421 & 1.366 & 1.256 & 0.737 & 0.455 & 1.619 & 1.513 \\
8B-OProj-MBS2 & 16384 & 512 & 4096 & 0.463 & 0.380 & 1.217 & 1.096 & 0.484 & 0.397 & 1.220 & 1.083 \\
405B-Down-MBS4 & 32768 & 6656 & 16384 & 9.128 & 7.253 & 1.258 & 1.260 & 10.103 & 8.679 & 1.164 & 1.181 \\
405B-OProj-MBS4 & 32768 & 2048 & 16384 & 3.983 & 2.788 & 1.429 & 1.415 & 4.554 & 3.665 & 1.243 & 1.205 \\
70B-Down-MBS4 & 32768 & 3584 & 8192 & 2.595 & 2.070 & 1.253 & 1.351 & 2.993 & 2.160 & 1.386 & 1.417 \\
70B-OProj-MBS4 & 32768 & 1024 & 8192 & 1.731 & 1.351 & 1.281 & 1.180 & 1.952 & 1.329 & 1.469 & 1.469 \\
8B-Down-MBS4 & 32768 & 1792 & 4096 & 1.112 & 0.736 & 1.512 & 1.423 & 1.295 & 0.843 & 1.536 & 1.515 \\
8B-OProj-MBS4 & 32768 & 512 & 4096 & 0.820 & 0.694 & 1.181 & 1.058 & 0.864 & 0.722 & 1.197 & 1.184 \\
405B-Down-MBS8 & 65536 & 6656 & 16384 & 18.304 & 14.027 & 1.305 & 1.274 & 20.140 & 17.485 & 1.152 & 1.205 \\
405B-OProj-MBS8 & 65536 & 2048 & 16384 & 8.486 & 5.205 & 1.630 & 1.629 & 9.607 & 7.415 & 1.296 & 1.316 \\
70B-Down-MBS8 & 65536 & 3584 & 8192 & 5.649 & 4.066 & 1.389 & 1.417 & 6.696 & 5.148 & 1.301 & 1.329 \\
70B-OProj-MBS8 & 65536 & 1024 & 8192 & 3.390 & 2.511 & 1.350 & 1.252 & 3.682 & 2.670 & 1.379 & 1.467 \\
8B-Down-MBS8 & 65536 & 1792 & 4096 & 2.049 & 1.370 & 1.496 & 1.501 & 2.418 & 1.670 & 1.448 & 1.551 \\
8B-OProj-MBS8 & 65536 & 512 & 4096 & 1.566 & 1.281 & 1.222 & 1.130 & 1.641 & 1.395 & 1.176 & 1.201 \\
\emph{geometric mean} & & & & & & \emph{1.305} & \emph{1.220} & & & \emph{1.357} & \emph{1.302} \\
\bottomrule
\end{tabular}
\end{footnotesize}
\end{center}
\vskip -0.1in
\end{table*}

\begin{table*}[p]
\caption{TP+SP microbenchmark, AG+GEMM, row dgrad, TP~8, MI355X: per iteration the slowest rank, median. Base: the RCCL collective and the GEMM run back to back on one stream with no overlap (BF16: torch.mm (hipBLASLt); MXFP8: a communication-free HipKittens MXFP8 GEMM). Speedup is base over \sysname{}; vs unf.\ is \sysname{}'s communication and GEMM WGs run back to back over the fused kernel.}
\label{tab:micro-tpsp-cfg3-MI355X}
\vskip 0.1in
\begin{center}
\begin{footnotesize}
\setlength{\tabcolsep}{3pt}
\begin{tabular}{lrrrrrrrrrrr}
\toprule
 & & & & \multicolumn{4}{c}{BF16} & \multicolumn{4}{c}{MXFP8} \\
\cmidrule(lr){5-8}\cmidrule(lr){9-12}
Shape & M & N & K & Base ms & \sysname{} ms & Speedup & vs unf. & Base ms & \sysname{} ms & Speedup & vs unf. \\
\midrule
405B-Down-MBS1 & 8192 & 6656 & 16384 & 2.229 & 1.919 & 1.162 & 1.220 & 2.438 & 1.782 & 1.368 & 1.267 \\
405B-OProj-MBS1 & 8192 & 2048 & 16384 & 1.152 & 0.966 & 1.192 & 1.124 & 1.233 & 0.907 & 1.360 & 1.310 \\
70B-Down-MBS1 & 8192 & 3584 & 8192 & 0.782 & 0.630 & 1.241 & 1.209 & 0.827 & 0.569 & 1.453 & 1.462 \\
70B-OProj-MBS1 & 8192 & 1024 & 8192 & 0.516 & 0.435 & 1.184 & 1.096 & 0.667 & 0.396 & 1.684 & 1.260 \\
8B-Down-MBS1 & 8192 & 1792 & 4096 & 0.396 & 0.279 & 1.416 & 1.086 & 0.457 & 0.262 & 1.746 & 1.457 \\
8B-OProj-MBS1 & 8192 & 512 & 4096 & 0.328 & 0.226 & 1.450 & 1.073 & 0.311 & 0.248 & 1.255 & 1.101 \\
405B-Down-MBS2 & 16384 & 6656 & 16384 & 4.190 & 3.409 & 1.229 & 1.268 & 4.806 & 3.604 & 1.333 & 1.268 \\
405B-OProj-MBS2 & 16384 & 2048 & 16384 & 2.238 & 1.622 & 1.380 & 1.336 & 2.403 & 1.858 & 1.293 & 1.305 \\
70B-Down-MBS2 & 16384 & 3584 & 8192 & 1.505 & 1.119 & 1.345 & 1.352 & 1.615 & 1.046 & 1.544 & 1.519 \\
70B-OProj-MBS2 & 16384 & 1024 & 8192 & 0.931 & 0.804 & 1.157 & 1.092 & 1.101 & 0.720 & 1.528 & 1.445 \\
8B-Down-MBS2 & 16384 & 1792 & 4096 & 0.596 & 0.438 & 1.360 & 1.299 & 0.732 & 0.445 & 1.646 & 1.565 \\
8B-OProj-MBS2 & 16384 & 512 & 4096 & 0.458 & 0.401 & 1.140 & 1.070 & 0.499 & 0.414 & 1.206 & 1.092 \\
405B-Down-MBS4 & 32768 & 6656 & 16384 & 8.425 & 6.467 & 1.303 & 1.307 & 9.360 & 7.264 & 1.289 & 1.286 \\
405B-OProj-MBS4 & 32768 & 2048 & 16384 & 4.366 & 2.924 & 1.493 & 1.465 & 4.724 & 3.674 & 1.286 & 1.276 \\
70B-Down-MBS4 & 32768 & 3584 & 8192 & 2.910 & 2.067 & 1.408 & 1.390 & 3.157 & 1.981 & 1.594 & 1.595 \\
70B-OProj-MBS4 & 32768 & 1024 & 8192 & 1.815 & 1.433 & 1.267 & 1.209 & 2.023 & 1.420 & 1.424 & 1.464 \\
8B-Down-MBS4 & 32768 & 1792 & 4096 & 1.137 & 0.766 & 1.485 & 1.455 & 1.304 & 0.828 & 1.574 & 1.584 \\
8B-OProj-MBS4 & 32768 & 512 & 4096 & 0.835 & 0.741 & 1.128 & 1.059 & 0.878 & 0.759 & 1.156 & 1.171 \\
405B-Down-MBS8 & 65536 & 6656 & 16384 & 16.544 & 12.430 & 1.331 & 1.338 & 18.272 & 14.587 & 1.253 & 1.298 \\
405B-OProj-MBS8 & 65536 & 2048 & 16384 & 8.603 & 5.359 & 1.605 & 1.577 & 9.597 & 7.376 & 1.301 & 1.318 \\
70B-Down-MBS8 & 65536 & 3584 & 8192 & 5.655 & 3.903 & 1.449 & 1.443 & 6.479 & 4.341 & 1.492 & 1.500 \\
70B-OProj-MBS8 & 65536 & 1024 & 8192 & 3.516 & 2.675 & 1.314 & 1.281 & 3.833 & 2.846 & 1.347 & 1.470 \\
8B-Down-MBS8 & 65536 & 1792 & 4096 & 2.162 & 1.453 & 1.487 & 1.464 & 2.454 & 1.646 & 1.491 & 1.655 \\
8B-OProj-MBS8 & 65536 & 512 & 4096 & 1.613 & 1.370 & 1.177 & 1.115 & 1.687 & 1.472 & 1.146 & 1.197 \\
\emph{geometric mean} & & & & & & \emph{1.314} & \emph{1.255} & & & \emph{1.398} & \emph{1.360} \\
\bottomrule
\end{tabular}
\end{footnotesize}
\end{center}
\vskip -0.1in
\end{table*}

\clearpage

\begin{table*}[p]
\caption{TP+SP microbenchmark, bulk AG, column dgrad, TP~8, MI350X: per iteration the slowest rank, median. Base: the RCCL collective and the GEMM run back to back on one stream with no overlap (BF16: torch.mm (hipBLASLt); MXFP8: a communication-free HipKittens MXFP8 GEMM). Speedup is base over \sysname{}; vs unf.\ is \sysname{}'s communication and GEMM WGs run back to back over the fused kernel.}
\label{tab:micro-tpsp-cfg5-MI350X}
\vskip 0.1in
\begin{center}
\begin{footnotesize}
\setlength{\tabcolsep}{3pt}
\begin{tabular}{lrrrrrrrrrrr}
\toprule
 & & & & \multicolumn{4}{c}{BF16} & \multicolumn{4}{c}{MXFP8} \\
\cmidrule(lr){5-8}\cmidrule(lr){9-12}
Shape & M & N & K & Base ms & \sysname{} ms & Speedup & vs unf. & Base ms & \sysname{} ms & Speedup & vs unf. \\
\midrule
405B-FC1-MBS1 & 8192 & 16384 & 13312 & 3.794 & 3.386 & 1.121 & 1.257 & 1.813 & 1.639 & 1.106 & 1.145 \\
405B-QKV-MBS1 & 8192 & 16384 & 2304 & 1.178 & 0.669 & 1.761 & 1.743 & 0.665 & 0.391 & 1.701 & 1.722 \\
70B-FC1-MBS1 & 8192 & 8192 & 7168 & 1.055 & 0.879 & 1.200 & 1.606 & 0.598 & 0.478 & 1.251 & 1.355 \\
70B-QKV-MBS1 & 8192 & 8192 & 1280 & 0.548 & 0.324 & 1.694 & 1.582 & 0.319 & 0.197 & 1.621 & 1.631 \\
8B-FC1-MBS1 & 8192 & 4096 & 3584 & 0.493 & 0.254 & 1.942 & 1.511 & 0.253 & 0.173 & 1.468 & 1.583 \\
8B-QKV-MBS1 & 8192 & 4096 & 768 & 0.364 & 0.176 & 2.062 & 1.342 & 0.185 & 0.121 & 1.524 & 1.419 \\
405B-FC1-MBS2 & 16384 & 16384 & 13312 & 8.014 & 6.959 & 1.152 & 1.354 & 4.015 & 3.670 & 1.094 & 1.133 \\
405B-QKV-MBS2 & 16384 & 16384 & 2304 & 2.232 & 1.393 & 1.603 & 1.647 & 1.279 & 0.751 & 1.702 & 1.770 \\
70B-FC1-MBS2 & 16384 & 8192 & 7168 & 1.991 & 1.815 & 1.097 & 1.577 & 1.139 & 0.904 & 1.260 & 1.380 \\
70B-QKV-MBS2 & 16384 & 8192 & 1280 & 1.008 & 0.602 & 1.676 & 1.578 & 0.576 & 0.355 & 1.620 & 1.547 \\
8B-FC1-MBS2 & 16384 & 4096 & 3584 & 0.753 & 0.493 & 1.529 & 1.569 & 0.430 & 0.286 & 1.500 & 1.698 \\
8B-QKV-MBS2 & 16384 & 4096 & 768 & 0.499 & 0.323 & 1.546 & 1.388 & 0.296 & 0.205 & 1.443 & 1.498 \\
405B-FC1-MBS4 & 32768 & 16384 & 13312 & 16.086 & 13.411 & 1.199 & 1.374 & 8.535 & 7.590 & 1.125 & 1.183 \\
405B-QKV-MBS4 & 32768 & 16384 & 2304 & 4.425 & 2.824 & 1.567 & 1.660 & 2.552 & 1.478 & 1.727 & 1.811 \\
70B-FC1-MBS4 & 32768 & 8192 & 7168 & 4.335 & 3.719 & 1.166 & 1.639 & 2.289 & 1.666 & 1.374 & 1.537 \\
70B-QKV-MBS4 & 32768 & 8192 & 1280 & 1.920 & 1.218 & 1.576 & 1.556 & 1.094 & 0.710 & 1.542 & 1.553 \\
8B-FC1-MBS4 & 32768 & 4096 & 3584 & 1.382 & 0.933 & 1.480 & 1.720 & 0.797 & 0.533 & 1.495 & 1.675 \\
8B-QKV-MBS4 & 32768 & 4096 & 768 & 0.908 & 0.617 & 1.472 & 1.403 & 0.531 & 0.385 & 1.378 & 1.357 \\
405B-FC1-MBS8 & 65536 & 16384 & 13312 & 31.460 & 26.567 & 1.184 & 1.390 & 17.226 & 14.820 & 1.162 & 1.221 \\
405B-QKV-MBS8 & 65536 & 16384 & 2304 & 9.445 & 5.904 & 1.600 & 1.680 & 5.114 & 3.031 & 1.687 & 1.773 \\
70B-FC1-MBS8 & 65536 & 8192 & 7168 & 9.304 & 7.733 & 1.203 & 1.636 & 4.825 & 3.955 & 1.220 & 1.330 \\
70B-QKV-MBS8 & 65536 & 8192 & 1280 & 3.791 & 2.456 & 1.544 & 1.569 & 2.164 & 1.445 & 1.498 & 1.521 \\
8B-FC1-MBS8 & 65536 & 4096 & 3584 & 2.661 & 1.918 & 1.387 & 1.638 & 1.549 & 1.076 & 1.440 & 1.649 \\
8B-QKV-MBS8 & 65536 & 4096 & 768 & 1.751 & 1.242 & 1.410 & 1.386 & 1.002 & 0.754 & 1.328 & 1.364 \\
\emph{geometric mean} & & & & & & \emph{1.443} & \emph{1.528} & & & \emph{1.414} & \emph{1.480} \\
\bottomrule
\end{tabular}
\end{footnotesize}
\end{center}
\vskip -0.1in
\end{table*}

\begin{table*}[p]
\caption{TP+SP microbenchmark, bulk AG, column dgrad, TP~8, MI355X: per iteration the slowest rank, median. Base: the RCCL collective and the GEMM run back to back on one stream with no overlap (BF16: torch.mm (hipBLASLt); MXFP8: a communication-free HipKittens MXFP8 GEMM). Speedup is base over \sysname{}; vs unf.\ is \sysname{}'s communication and GEMM WGs run back to back over the fused kernel.}
\label{tab:micro-tpsp-cfg5-MI355X}
\vskip 0.1in
\begin{center}
\begin{footnotesize}
\setlength{\tabcolsep}{3pt}
\begin{tabular}{lrrrrrrrrrrr}
\toprule
 & & & & \multicolumn{4}{c}{BF16} & \multicolumn{4}{c}{MXFP8} \\
\cmidrule(lr){5-8}\cmidrule(lr){9-12}
Shape & M & N & K & Base ms & \sysname{} ms & Speedup & vs unf. & Base ms & \sysname{} ms & Speedup & vs unf. \\
\midrule
405B-FC1-MBS1 & 8192 & 16384 & 13312 & 3.519 & 3.047 & 1.155 & 1.345 & 1.937 & 1.629 & 1.189 & 1.232 \\
405B-QKV-MBS1 & 8192 & 16384 & 2304 & 1.246 & 0.673 & 1.850 & 1.776 & 0.689 & 0.401 & 1.718 & 1.748 \\
70B-FC1-MBS1 & 8192 & 8192 & 7168 & 1.168 & 0.897 & 1.302 & 1.619 & 0.647 & 0.506 & 1.279 & 1.376 \\
70B-QKV-MBS1 & 8192 & 8192 & 1280 & 0.558 & 0.339 & 1.645 & 1.511 & 0.320 & 0.203 & 1.577 & 1.581 \\
8B-FC1-MBS1 & 8192 & 4096 & 3584 & 0.493 & 0.269 & 1.838 & 1.498 & 0.252 & 0.169 & 1.495 & 1.570 \\
8B-QKV-MBS1 & 8192 & 4096 & 768 & 0.347 & 0.187 & 1.850 & 1.336 & 0.180 & 0.127 & 1.423 & 1.405 \\
405B-FC1-MBS2 & 16384 & 16384 & 13312 & 7.095 & 6.006 & 1.181 & 1.405 & 3.800 & 3.120 & 1.218 & 1.242 \\
405B-QKV-MBS2 & 16384 & 16384 & 2304 & 2.406 & 1.394 & 1.727 & 1.725 & 1.341 & 0.792 & 1.693 & 1.745 \\
70B-FC1-MBS2 & 16384 & 8192 & 7168 & 2.264 & 1.709 & 1.325 & 1.689 & 1.238 & 0.911 & 1.358 & 1.467 \\
70B-QKV-MBS2 & 16384 & 8192 & 1280 & 1.042 & 0.646 & 1.613 & 1.518 & 0.584 & 0.374 & 1.560 & 1.517 \\
8B-FC1-MBS2 & 16384 & 4096 & 3584 & 0.792 & 0.489 & 1.620 & 1.625 & 0.446 & 0.286 & 1.556 & 1.792 \\
8B-QKV-MBS2 & 16384 & 4096 & 768 & 0.502 & 0.344 & 1.459 & 1.344 & 0.294 & 0.212 & 1.388 & 1.476 \\
405B-FC1-MBS4 & 32768 & 16384 & 13312 & 13.862 & 11.584 & 1.197 & 1.421 & 7.633 & 6.282 & 1.215 & 1.253 \\
405B-QKV-MBS4 & 32768 & 16384 & 2304 & 4.701 & 2.777 & 1.693 & 1.732 & 2.651 & 1.561 & 1.699 & 1.762 \\
70B-FC1-MBS4 & 32768 & 8192 & 7168 & 4.385 & 3.456 & 1.269 & 1.651 & 2.411 & 1.758 & 1.371 & 1.463 \\
70B-QKV-MBS4 & 32768 & 8192 & 1280 & 2.000 & 1.292 & 1.548 & 1.521 & 1.126 & 0.712 & 1.581 & 1.582 \\
8B-FC1-MBS4 & 32768 & 4096 & 3584 & 1.504 & 0.928 & 1.621 & 1.725 & 0.840 & 0.521 & 1.613 & 1.812 \\
8B-QKV-MBS4 & 32768 & 4096 & 768 & 0.935 & 0.662 & 1.411 & 1.344 & 0.532 & 0.403 & 1.320 & 1.345 \\
405B-FC1-MBS8 & 65536 & 16384 & 13312 & 26.956 & 22.603 & 1.193 & 1.460 & 14.849 & 12.191 & 1.218 & 1.282 \\
405B-QKV-MBS8 & 65536 & 16384 & 2304 & 9.374 & 5.497 & 1.705 & 1.742 & 5.235 & 3.045 & 1.719 & 1.775 \\
70B-FC1-MBS8 & 65536 & 8192 & 7168 & 8.699 & 6.880 & 1.264 & 1.667 & 4.793 & 3.504 & 1.368 & 1.461 \\
70B-QKV-MBS8 & 65536 & 8192 & 1280 & 3.880 & 2.596 & 1.495 & 1.568 & 2.228 & 1.465 & 1.521 & 1.545 \\
8B-FC1-MBS8 & 65536 & 4096 & 3584 & 2.950 & 1.875 & 1.573 & 1.685 & 1.641 & 1.070 & 1.533 & 1.721 \\
8B-QKV-MBS8 & 65536 & 4096 & 768 & 1.787 & 1.318 & 1.356 & 1.355 & 1.025 & 0.764 & 1.342 & 1.375 \\
\emph{geometric mean} & & & & & & \emph{1.479} & \emph{1.546} & & & \emph{1.447} & \emph{1.511} \\
\bottomrule
\end{tabular}
\end{footnotesize}
\end{center}
\vskip -0.1in
\end{table*}

\begin{table*}[p]
\caption{TP+SP microbenchmark, bulk RS, column wgrad, TP~8, MI350X: per iteration the slowest rank, median. Base: the RCCL collective and the GEMM run back to back on one stream with no overlap (BF16: torch.mm (hipBLASLt); MXFP8: a communication-free HipKittens MXFP8 GEMM). Speedup is base over \sysname{}; vs unf.\ is \sysname{}'s communication and GEMM WGs run back to back over the fused kernel.}
\label{tab:micro-tpsp-cfg6-MI350X}
\vskip 0.1in
\begin{center}
\begin{footnotesize}
\setlength{\tabcolsep}{3pt}
\begin{tabular}{lrrrrrrrrrrr}
\toprule
 & & & & \multicolumn{4}{c}{BF16} & \multicolumn{4}{c}{MXFP8} \\
\cmidrule(lr){5-8}\cmidrule(lr){9-12}
Shape & M & N & K & Base ms & \sysname{} ms & Speedup & vs unf. & Base ms & \sysname{} ms & Speedup & vs unf. \\
\midrule
405B-FC1-MBS1 & 13312 & 16384 & 8192 & 3.971 & 3.256 & 1.220 & 1.589 & 2.262 & 1.653 & 1.369 & 1.505 \\
405B-QKV-MBS1 & 2304 & 16384 & 8192 & 1.145 & 0.660 & 1.735 & 1.793 & 1.011 & 0.625 & 1.619 & 1.516 \\
70B-FC1-MBS1 & 7168 & 8192 & 8192 & 1.242 & 0.771 & 1.611 & 1.610 & 0.825 & 0.552 & 1.496 & 1.472 \\
70B-QKV-MBS1 & 1280 & 8192 & 8192 & 0.559 & 0.336 & 1.663 & 1.558 & 0.479 & 0.333 & 1.436 & 1.328 \\
8B-FC1-MBS1 & 3584 & 4096 & 8192 & 0.444 & 0.238 & 1.864 & 1.870 & 0.321 & 0.183 & 1.752 & 1.670 \\
8B-QKV-MBS1 & 768 & 4096 & 8192 & 0.330 & 0.190 & 1.730 & 1.509 & 0.299 & 0.179 & 1.674 & 1.544 \\
405B-FC1-MBS2 & 13312 & 16384 & 16384 & 8.329 & 7.155 & 1.164 & 1.443 & 4.827 & 3.696 & 1.306 & 1.453 \\
405B-QKV-MBS2 & 2304 & 16384 & 16384 & 2.151 & 1.279 & 1.682 & 1.821 & 1.949 & 1.193 & 1.633 & 1.525 \\
70B-FC1-MBS2 & 7168 & 8192 & 16384 & 2.386 & 1.513 & 1.577 & 1.681 & 1.567 & 1.052 & 1.489 & 1.431 \\
70B-QKV-MBS2 & 1280 & 8192 & 16384 & 1.014 & 0.619 & 1.638 & 1.585 & 0.886 & 0.613 & 1.447 & 1.329 \\
8B-FC1-MBS2 & 3584 & 4096 & 16384 & 0.763 & 0.494 & 1.545 & 1.722 & 0.590 & 0.335 & 1.760 & 1.644 \\
8B-QKV-MBS2 & 768 & 4096 & 16384 & 0.525 & 0.364 & 1.441 & 1.249 & 0.552 & 0.325 & 1.699 & 1.568 \\
405B-FC1-MBS4 & 13312 & 16384 & 32768 & 16.695 & 14.465 & 1.154 & 1.453 & 10.298 & 7.812 & 1.318 & 1.383 \\
405B-QKV-MBS4 & 2304 & 16384 & 32768 & 4.193 & 2.573 & 1.630 & 1.885 & 3.780 & 2.339 & 1.616 & 1.524 \\
70B-FC1-MBS4 & 7168 & 8192 & 32768 & 4.731 & 3.673 & 1.288 & 1.488 & 3.151 & 2.129 & 1.480 & 1.404 \\
70B-QKV-MBS4 & 1280 & 8192 & 32768 & 1.922 & 1.195 & 1.608 & 1.602 & 1.713 & 1.174 & 1.458 & 1.336 \\
8B-FC1-MBS4 & 3584 & 4096 & 32768 & 1.396 & 0.967 & 1.444 & 1.759 & 1.115 & 0.622 & 1.792 & 1.675 \\
8B-QKV-MBS4 & 768 & 4096 & 32768 & 0.932 & 0.651 & 1.433 & 1.276 & 1.037 & 0.608 & 1.706 & 1.590 \\
405B-FC1-MBS8 & 13312 & 16384 & 65536 & 32.763 & 29.801 & 1.099 & 1.419 & 20.770 & 16.225 & 1.280 & 1.346 \\
405B-QKV-MBS8 & 2304 & 16384 & 65536 & 9.099 & 5.821 & 1.563 & 1.774 & 7.988 & 4.630 & 1.725 & 1.551 \\
70B-FC1-MBS8 & 7168 & 8192 & 65536 & 10.576 & 8.094 & 1.307 & 1.417 & 6.813 & 4.902 & 1.390 & 1.326 \\
70B-QKV-MBS8 & 1280 & 8192 & 65536 & 3.724 & 2.384 & 1.562 & 1.613 & 3.315 & 2.294 & 1.445 & 1.339 \\
8B-FC1-MBS8 & 3584 & 4096 & 65536 & 2.695 & 1.956 & 1.378 & 1.798 & 2.188 & 1.207 & 1.813 & 1.703 \\
8B-QKV-MBS8 & 768 & 4096 & 65536 & 1.736 & 1.226 & 1.416 & 1.302 & 2.017 & 1.162 & 1.735 & 1.611 \\
\emph{geometric mean} & & & & & & \emph{1.476} & \emph{1.582} & & & \emph{1.551} & \emph{1.486} \\
\bottomrule
\end{tabular}
\end{footnotesize}
\end{center}
\vskip -0.1in
\end{table*}

\begin{table*}[p]
\caption{TP+SP microbenchmark, bulk RS, column wgrad, TP~8, MI355X: per iteration the slowest rank, median. Base: the RCCL collective and the GEMM run back to back on one stream with no overlap (BF16: torch.mm (hipBLASLt); MXFP8: a communication-free HipKittens MXFP8 GEMM). Speedup is base over \sysname{}; vs unf.\ is \sysname{}'s communication and GEMM WGs run back to back over the fused kernel.}
\label{tab:micro-tpsp-cfg6-MI355X}
\vskip 0.1in
\begin{center}
\begin{footnotesize}
\setlength{\tabcolsep}{3pt}
\begin{tabular}{lrrrrrrrrrrr}
\toprule
 & & & & \multicolumn{4}{c}{BF16} & \multicolumn{4}{c}{MXFP8} \\
\cmidrule(lr){5-8}\cmidrule(lr){9-12}
Shape & M & N & K & Base ms & \sysname{} ms & Speedup & vs unf. & Base ms & \sysname{} ms & Speedup & vs unf. \\
\midrule
405B-FC1-MBS1 & 13312 & 16384 & 8192 & 3.620 & 3.015 & 1.201 & 1.586 & 2.323 & 1.612 & 1.441 & 1.602 \\
405B-QKV-MBS1 & 2304 & 16384 & 8192 & 1.226 & 0.691 & 1.774 & 1.925 & 1.065 & 0.667 & 1.598 & 1.501 \\
70B-FC1-MBS1 & 7168 & 8192 & 8192 & 1.228 & 0.841 & 1.460 & 1.637 & 0.852 & 0.533 & 1.598 & 1.581 \\
70B-QKV-MBS1 & 1280 & 8192 & 8192 & 0.566 & 0.357 & 1.583 & 1.518 & 0.490 & 0.356 & 1.375 & 1.289 \\
8B-FC1-MBS1 & 3584 & 4096 & 8192 & 0.446 & 0.234 & 1.907 & 1.926 & 0.323 & 0.194 & 1.664 & 1.594 \\
8B-QKV-MBS1 & 768 & 4096 & 8192 & 0.304 & 0.206 & 1.471 & 1.432 & 0.298 & 0.194 & 1.536 & 1.472 \\
405B-FC1-MBS2 & 13312 & 16384 & 16384 & 7.231 & 6.188 & 1.169 & 1.513 & 4.638 & 3.348 & 1.385 & 1.491 \\
405B-QKV-MBS2 & 2304 & 16384 & 16384 & 2.377 & 1.361 & 1.747 & 1.914 & 2.058 & 1.276 & 1.613 & 1.512 \\
70B-FC1-MBS2 & 7168 & 8192 & 16384 & 2.329 & 1.671 & 1.393 & 1.613 & 1.622 & 1.008 & 1.610 & 1.560 \\
70B-QKV-MBS2 & 1280 & 8192 & 16384 & 1.042 & 0.659 & 1.582 & 1.552 & 0.918 & 0.658 & 1.394 & 1.291 \\
8B-FC1-MBS2 & 3584 & 4096 & 16384 & 0.805 & 0.505 & 1.594 & 1.819 & 0.597 & 0.354 & 1.686 & 1.617 \\
8B-QKV-MBS2 & 768 & 4096 & 16384 & 0.509 & 0.383 & 1.328 & 1.277 & 0.556 & 0.354 & 1.570 & 1.499 \\
405B-FC1-MBS4 & 13312 & 16384 & 32768 & 14.160 & 12.013 & 1.179 & 1.508 & 9.387 & 6.753 & 1.390 & 1.463 \\
405B-QKV-MBS4 & 2304 & 16384 & 32768 & 4.596 & 2.625 & 1.751 & 1.902 & 4.026 & 2.484 & 1.621 & 1.528 \\
70B-FC1-MBS4 & 7168 & 8192 & 32768 & 4.537 & 3.368 & 1.347 & 1.586 & 3.215 & 1.983 & 1.621 & 1.558 \\
70B-QKV-MBS4 & 1280 & 8192 & 32768 & 2.010 & 1.278 & 1.573 & 1.563 & 1.778 & 1.262 & 1.409 & 1.298 \\
8B-FC1-MBS4 & 3584 & 4096 & 32768 & 1.546 & 0.993 & 1.556 & 1.848 & 1.137 & 0.660 & 1.723 & 1.635 \\
8B-QKV-MBS4 & 768 & 4096 & 32768 & 0.926 & 0.690 & 1.341 & 1.302 & 1.056 & 0.656 & 1.610 & 1.512 \\
405B-FC1-MBS8 & 13312 & 16384 & 65536 & 27.920 & 23.735 & 1.176 & 1.515 & 18.367 & 13.417 & 1.369 & 1.435 \\
405B-QKV-MBS8 & 2304 & 16384 & 65536 & 9.101 & 5.212 & 1.746 & 1.943 & 8.139 & 4.912 & 1.657 & 1.554 \\
70B-FC1-MBS8 & 7168 & 8192 & 65536 & 9.363 & 6.846 & 1.368 & 1.520 & 6.592 & 4.424 & 1.490 & 1.425 \\
70B-QKV-MBS8 & 1280 & 8192 & 65536 & 3.918 & 2.509 & 1.562 & 1.562 & 3.466 & 2.478 & 1.399 & 1.305 \\
8B-FC1-MBS8 & 3584 & 4096 & 65536 & 3.025 & 1.967 & 1.538 & 1.836 & 2.231 & 1.278 & 1.745 & 1.653 \\
8B-QKV-MBS8 & 768 & 4096 & 65536 & 1.764 & 1.301 & 1.355 & 1.326 & 2.057 & 1.262 & 1.629 & 1.524 \\
\emph{geometric mean} & & & & & & \emph{1.474} & \emph{1.617} & & & \emph{1.543} & \emph{1.492} \\
\bottomrule
\end{tabular}
\end{footnotesize}
\end{center}
\vskip -0.1in
\end{table*}

\clearpage

\begin{table*}[p]
\caption{MoE microbenchmark, BF16, forward pass, EP~8: per call the slowest rank, median. Baseline: RCCL A2A with TE's grouped GEMMs. Speedup is baseline over \sysname{}; vs \sysname{} serial is the same \sysname{} kernels as separate launches over the fused layer.}
\label{tab:micro-moe-bf16-fwd}
\vskip 0.1in
\begin{center}
\begin{footnotesize}
\setlength{\tabcolsep}{4pt}
\begin{tabular}{llrrrrr}
\toprule
Routing & Model & Baseline ms & \sysname{} ms & \sysname{} TFLOPS & Speedup & vs \sysname{} serial \\
\midrule
\multicolumn{7}{l}{\textbf{MI350X}} \\
balanced & DeepSeek-V3 & 17.041 & 7.670 & 752.6 & 2.222 & 1.463 \\
 & Qwen3.5-397B & 11.026 & 3.791 & 543.8 & 2.908 & 1.610 \\
 & DeepSeek-V4-Pro & 17.443 & 8.531 & 761.2 & 2.045 & 1.314 \\
 & Kimi-K2.7-Code & 18.219 & 8.117 & 711.1 & 2.244 & 1.438 \\
 & GLM-5.2 & 14.753 & 6.562 & 754.0 & 2.248 & 1.476 \\
 & Qwen3-235B & 8.742 & 3.405 & 726.6 & 2.568 & 1.568 \\
 & Qwen3-30B & 4.158 & 1.555 & 397.7 & 2.674 & 1.508 \\
 & Llama 4 Scout & 3.358 & 2.381 & 865.8 & 1.410 & 1.099 \\
 & Llama 4 Maverick & 4.373 & 2.710 & 760.8 & 1.614 & 1.212 \\
 & \emph{geometric mean} & & & & \emph{2.164} & \emph{1.400} \\
skewed & DeepSeek-V3 & 31.357 & 15.792 & 365.5 & 1.986 & 1.324 \\
 & Qwen3.5-397B & 19.649 & 7.848 & 262.7 & 2.504 & 1.402 \\
 & DeepSeek-V4-Pro & 30.492 & 17.083 & 380.1 & 1.785 & 1.286 \\
 & Kimi-K2.7-Code & 32.022 & 16.005 & 360.7 & 2.001 & 1.335 \\
 & GLM-5.2 & 27.324 & 13.640 & 362.7 & 2.003 & 1.333 \\
 & Qwen3-235B & 16.608 & 7.547 & 327.8 & 2.201 & 1.373 \\
 & Qwen3-30B & 7.093 & 3.126 & 197.9 & 2.269 & 1.216 \\
 & Llama 4 Scout & 7.586 & 5.737 & 359.4 & 1.322 & 1.078 \\
 & Llama 4 Maverick & 7.679 & 5.651 & 364.8 & 1.359 & 1.223 \\
 & \emph{geometric mean} & & & & \emph{1.898} & \emph{1.282} \\
\midrule
\multicolumn{7}{l}{\textbf{MI355X}} \\
balanced & DeepSeek-V3 & 15.856 & 6.806 & 848.1 & 2.330 & 1.592 \\
 & Qwen3.5-397B & 11.215 & 4.008 & 514.4 & 2.799 & 1.570 \\
 & DeepSeek-V4-Pro & 16.249 & 7.091 & 915.8 & 2.292 & 1.479 \\
 & Kimi-K2.7-Code & 17.165 & 7.081 & 815.2 & 2.424 & 1.571 \\
 & GLM-5.2 & 13.696 & 5.882 & 841.2 & 2.329 & 1.591 \\
 & Qwen3-235B & 8.421 & 3.463 & 714.4 & 2.432 & 1.601 \\
 & Qwen3-30B & 4.103 & 1.644 & 376.1 & 2.495 & 1.456 \\
 & Llama 4 Scout & 3.183 & 2.287 & 901.6 & 1.392 & 1.158 \\
 & Llama 4 Maverick & 4.090 & 2.423 & 850.8 & 1.688 & 1.277 \\
 & \emph{geometric mean} & & & & \emph{2.199} & \emph{1.469} \\
skewed & DeepSeek-V3 & 30.078 & 14.522 & 397.5 & 2.071 & 1.379 \\
 & Qwen3.5-397B & 19.235 & 8.481 & 243.1 & 2.268 & 1.314 \\
 & DeepSeek-V4-Pro & 29.068 & 14.683 & 442.3 & 1.980 & 1.394 \\
 & Kimi-K2.7-Code & 30.608 & 14.620 & 394.8 & 2.094 & 1.401 \\
 & GLM-5.2 & 26.230 & 12.631 & 391.7 & 2.077 & 1.376 \\
 & Qwen3-235B & 15.958 & 7.338 & 337.1 & 2.175 & 1.388 \\
 & Qwen3-30B & 7.048 & 3.368 & 183.6 & 2.093 & 1.198 \\
 & Llama 4 Scout & 6.574 & 5.123 & 402.4 & 1.283 & 1.115 \\
 & Llama 4 Maverick & 7.143 & 4.966 & 415.2 & 1.438 & 1.241 \\
 & \emph{geometric mean} & & & & \emph{1.911} & \emph{1.308} \\
\bottomrule
\end{tabular}
\end{footnotesize}
\end{center}
\vskip -0.1in
\end{table*}

\begin{table*}[p]
\caption{MoE microbenchmark, BF16, backward pass, EP~8: per call the slowest rank, median. Baseline: RCCL A2A with TE's grouped GEMMs. Speedup is baseline over \sysname{}; vs \sysname{} serial is the same \sysname{} kernels as separate launches over the fused layer.}
\label{tab:micro-moe-bf16-bwd}
\vskip 0.1in
\begin{center}
\begin{footnotesize}
\setlength{\tabcolsep}{4pt}
\begin{tabular}{llrrrrr}
\toprule
Routing & Model & Baseline ms & \sysname{} ms & \sysname{} TFLOPS & Speedup & vs \sysname{} serial \\
\midrule
\multicolumn{7}{l}{\textbf{MI350X}} \\
balanced & DeepSeek-V3 & 25.939 & 14.337 & 805.2 & 1.809 & 1.237 \\
 & Qwen3.5-397B & 16.245 & 5.918 & 696.7 & 2.745 & 1.479 \\
 & DeepSeek-V4-Pro & 29.207 & 16.807 & 772.8 & 1.738 & 1.187 \\
 & Kimi-K2.7-Code & 26.821 & 14.994 & 770.0 & 1.789 & 1.249 \\
 & GLM-5.2 & 23.205 & 12.294 & 804.9 & 1.888 & 1.252 \\
 & Qwen3-235B & 13.479 & 6.040 & 819.1 & 2.232 & 1.408 \\
 & Qwen3-30B & 6.068 & 1.793 & 690.0 & 3.385 & 1.703 \\
 & Llama 4 Scout & 6.675 & 4.383 & 940.8 & 1.523 & 1.111 \\
 & Llama 4 Maverick & 9.309 & 6.198 & 665.2 & 1.502 & 1.113 \\
 & \emph{geometric mean} & & & & \emph{1.996} & \emph{1.293} \\
skewed & DeepSeek-V3 & 48.415 & 30.535 & 378.1 & 1.586 & 1.196 \\
 & Qwen3.5-397B & 28.235 & 11.970 & 344.4 & 2.359 & 1.467 \\
 & DeepSeek-V4-Pro & 57.860 & 34.486 & 376.6 & 1.678 & 1.127 \\
 & Kimi-K2.7-Code & 50.908 & 30.922 & 373.4 & 1.646 & 1.220 \\
 & GLM-5.2 & 42.471 & 26.627 & 371.6 & 1.595 & 1.186 \\
 & Qwen3-235B & 24.938 & 13.154 & 376.1 & 1.896 & 1.356 \\
 & Qwen3-30B & 10.382 & 4.152 & 297.9 & 2.501 & 1.574 \\
 & Llama 4 Scout & 14.418 & 10.122 & 407.3 & 1.424 & 1.101 \\
 & Llama 4 Maverick & 15.242 & 11.350 & 363.3 & 1.343 & 1.095 \\
 & \emph{geometric mean} & & & & \emph{1.744} & \emph{1.248} \\
\midrule
\multicolumn{7}{l}{\textbf{MI355X}} \\
balanced & DeepSeek-V3 & 24.104 & 11.493 & 1004.5 & 2.097 & 1.403 \\
 & Qwen3.5-397B & 15.765 & 5.162 & 798.8 & 3.054 & 1.647 \\
 & DeepSeek-V4-Pro & 27.974 & 13.883 & 935.6 & 2.015 & 1.270 \\
 & Kimi-K2.7-Code & 25.323 & 12.049 & 958.2 & 2.102 & 1.398 \\
 & GLM-5.2 & 21.445 & 9.930 & 996.6 & 2.160 & 1.410 \\
 & Qwen3-235B & 12.770 & 5.176 & 956.0 & 2.467 & 1.559 \\
 & Qwen3-30B & 5.834 & 1.777 & 696.0 & 3.283 & 1.763 \\
 & Llama 4 Scout & 5.651 & 3.888 & 1060.4 & 1.453 & 1.162 \\
 & Llama 4 Maverick & 8.226 & 5.327 & 774.1 & 1.544 & 1.140 \\
 & \emph{geometric mean} & & & & \emph{2.169} & \emph{1.403} \\
skewed & DeepSeek-V3 & 46.070 & 25.534 & 452.1 & 1.804 & 1.297 \\
 & Qwen3.5-397B & 27.346 & 10.560 & 390.5 & 2.590 & 1.578 \\
 & DeepSeek-V4-Pro & 55.182 & 28.365 & 457.9 & 1.945 & 1.228 \\
 & Kimi-K2.7-Code & 49.450 & 25.540 & 452.0 & 1.936 & 1.331 \\
 & GLM-5.2 & 40.206 & 21.712 & 455.8 & 1.852 & 1.325 \\
 & Qwen3-235B & 23.386 & 11.339 & 436.3 & 2.062 & 1.460 \\
 & Qwen3-30B & 10.072 & 3.772 & 328.0 & 2.670 & 1.702 \\
 & Llama 4 Scout & 11.729 & 8.816 & 467.7 & 1.330 & 1.152 \\
 & Llama 4 Maverick & 13.823 & 9.801 & 420.7 & 1.410 & 1.135 \\
 & \emph{geometric mean} & & & & \emph{1.909} & \emph{1.345} \\
\bottomrule
\end{tabular}
\end{footnotesize}
\end{center}
\vskip -0.1in
\end{table*}

\begin{table*}[p]
\caption{MoE microbenchmark, MXFP8, forward pass, EP~8: per call the slowest rank, median. Baseline: RCCL A2A around Transformer Engine MXFP8 grouped GEMMs. Speedup is baseline over \sysname{}; vs \sysname{} serial is the same \sysname{} kernels as separate launches over the fused layer.}
\label{tab:micro-moe-mxfp8-fwd}
\vskip 0.1in
\begin{center}
\begin{footnotesize}
\setlength{\tabcolsep}{4pt}
\begin{tabular}{llrrrrr}
\toprule
Routing & Model & TE MXFP8 ms & \sysname{} ms & \sysname{} TFLOPS & Speedup & vs \sysname{} serial \\
\midrule
\multicolumn{7}{l}{\textbf{MI350X}} \\
balanced & DeepSeek-V3 & 15.554 & 5.326 & 1083.9 & 2.921 & 1.498 \\
 & Qwen3.5-397B & 10.323 & 3.202 & 643.9 & 3.224 & 1.464 \\
 & DeepSeek-V4-Pro & 14.399 & 5.516 & 1177.4 & 2.611 & 1.424 \\
 & Kimi-K2.7-Code & 15.916 & 5.514 & 1046.9 & 2.887 & 1.505 \\
 & GLM-5.2 & 13.561 & 4.637 & 1067.1 & 2.925 & 1.483 \\
 & Qwen3-235B & 8.661 & 2.846 & 869.1 & 3.043 & 1.418 \\
 & Qwen3-30B & 4.285 & 1.591 & 388.6 & 2.693 & 1.421 \\
 & Llama 4 Scout & 2.731 & 1.749 & 1178.8 & 1.562 & 1.111 \\
 & Llama 4 Maverick & 3.450 & 1.899 & 1085.9 & 1.817 & 1.240 \\
 & \emph{geometric mean} & & & & \emph{2.566} & \emph{1.390} \\
skewed & DeepSeek-V3 & 29.825 & 11.790 & 489.6 & 2.530 & 1.375 \\
 & Qwen3.5-397B & 19.397 & 6.396 & 322.4 & 3.033 & 1.389 \\
 & DeepSeek-V4-Pro & 27.322 & 11.497 & 564.9 & 2.377 & 1.364 \\
 & Kimi-K2.7-Code & 30.105 & 11.811 & 488.7 & 2.549 & 1.394 \\
 & GLM-5.2 & 25.839 & 10.215 & 484.4 & 2.530 & 1.386 \\
 & Qwen3-235B & 16.442 & 6.002 & 412.2 & 2.739 & 1.416 \\
 & Qwen3-30B & 7.770 & 2.839 & 217.9 & 2.737 & 1.320 \\
 & Llama 4 Scout & 5.586 & 3.905 & 527.9 & 1.430 & 1.080 \\
 & Llama 4 Maverick & 6.052 & 3.794 & 543.3 & 1.595 & 1.211 \\
 & \emph{geometric mean} & & & & \emph{2.328} & \emph{1.322} \\
\midrule
\multicolumn{7}{l}{\textbf{MI355X}} \\
balanced & DeepSeek-V3 & 14.836 & 5.325 & 1084.0 & 2.786 & 1.489 \\
 & Qwen3.5-397B & 10.013 & 3.364 & 612.8 & 2.976 & 1.439 \\
 & DeepSeek-V4-Pro & 13.527 & 5.140 & 1263.5 & 2.632 & 1.476 \\
 & Kimi-K2.7-Code & 15.154 & 5.363 & 1076.4 & 2.826 & 1.515 \\
 & GLM-5.2 & 12.950 & 4.664 & 1060.9 & 2.777 & 1.477 \\
 & Qwen3-235B & 8.353 & 2.870 & 862.0 & 2.911 & 1.442 \\
 & Qwen3-30B & 4.174 & 1.687 & 366.7 & 2.474 & 1.388 \\
 & Llama 4 Scout & 2.713 & 1.698 & 1214.2 & 1.598 & 1.136 \\
 & Llama 4 Maverick & 3.163 & 1.838 & 1121.5 & 1.721 & 1.242 \\
 & \emph{geometric mean} & & & & \emph{2.468} & \emph{1.395} \\
skewed & DeepSeek-V3 & 28.632 & 11.643 & 495.8 & 2.459 & 1.380 \\
 & Qwen3.5-397B & 18.822 & 6.836 & 301.6 & 2.753 & 1.342 \\
 & DeepSeek-V4-Pro & 26.344 & 10.897 & 595.9 & 2.418 & 1.399 \\
 & Kimi-K2.7-Code & 28.936 & 11.726 & 492.3 & 2.468 & 1.390 \\
 & GLM-5.2 & 24.851 & 10.066 & 491.5 & 2.469 & 1.391 \\
 & Qwen3-235B & 15.819 & 6.031 & 410.2 & 2.623 & 1.406 \\
 & Qwen3-30B & 7.634 & 2.964 & 208.7 & 2.576 & 1.287 \\
 & Llama 4 Scout & 5.361 & 3.620 & 569.6 & 1.481 & 1.093 \\
 & Llama 4 Maverick & 5.682 & 3.455 & 596.7 & 1.645 & 1.256 \\
 & \emph{geometric mean} & & & & \emph{2.276} & \emph{1.323} \\
\bottomrule
\end{tabular}
\end{footnotesize}
\end{center}
\vskip -0.1in
\end{table*}

\begin{table*}[p]
\caption{MoE microbenchmark, MXFP8, backward pass, EP~8: per call the slowest rank, median. Baseline: RCCL A2A around Transformer Engine MXFP8 grouped GEMMs. Speedup is baseline over \sysname{}; vs \sysname{} serial is the same \sysname{} kernels as separate launches over the fused layer.}
\label{tab:micro-moe-mxfp8-bwd}
\vskip 0.1in
\begin{center}
\begin{footnotesize}
\setlength{\tabcolsep}{4pt}
\begin{tabular}{llrrrrr}
\toprule
Routing & Model & TE MXFP8 ms & \sysname{} ms & \sysname{} TFLOPS & Speedup & vs \sysname{} serial \\
\midrule
\multicolumn{7}{l}{\textbf{MI350X}} \\
balanced & DeepSeek-V3 & 23.684 & 10.519 & 1097.5 & 2.251 & 1.430 \\
 & Qwen3.5-397B & 14.637 & 5.149 & 800.8 & 2.843 & 1.631 \\
 & DeepSeek-V4-Pro & 25.560 & 12.512 & 1038.1 & 2.043 & 1.316 \\
 & Kimi-K2.7-Code & 25.133 & 11.021 & 1047.6 & 2.281 & 1.476 \\
 & GLM-5.2 & 20.706 & 8.911 & 1110.5 & 2.324 & 1.452 \\
 & Qwen3-235B & 12.346 & 4.681 & 1056.9 & 2.637 & 1.531 \\
 & Qwen3-30B & 5.557 & 1.699 & 728.0 & 3.270 & 1.508 \\
 & Llama 4 Scout & 5.197 & 2.704 & 1524.5 & 1.921 & 1.317 \\
 & Llama 4 Maverick & 8.289 & 4.477 & 920.9 & 1.851 & 1.290 \\
 & \emph{geometric mean} & & & & \emph{2.343} & \emph{1.435} \\
skewed & DeepSeek-V3 & 44.307 & 21.928 & 526.5 & 2.021 & 1.387 \\
 & Qwen3.5-397B & 26.233 & 9.945 & 414.6 & 2.638 & 1.691 \\
 & DeepSeek-V4-Pro & 45.664 & 23.454 & 553.8 & 1.947 & 1.355 \\
 & Kimi-K2.7-Code & 45.325 & 21.599 & 534.5 & 2.098 & 1.472 \\
 & GLM-5.2 & 38.530 & 18.280 & 541.3 & 2.108 & 1.434 \\
 & Qwen3-235B & 23.169 & 10.053 & 492.2 & 2.305 & 1.520 \\
 & Qwen3-30B & 10.217 & 3.659 & 338.1 & 2.792 & 1.442 \\
 & Llama 4 Scout & 10.547 & 6.415 & 642.7 & 1.644 & 1.225 \\
 & Llama 4 Maverick & 12.913 & 7.873 & 523.7 & 1.640 & 1.172 \\
 & \emph{geometric mean} & & & & \emph{2.101} & \emph{1.403} \\
\midrule
\multicolumn{7}{l}{\textbf{MI355X}} \\
balanced & DeepSeek-V3 & 21.941 & 9.032 & 1278.2 & 2.429 & 1.586 \\
 & Qwen3.5-397B & 13.748 & 5.246 & 786.0 & 2.621 & 1.599 \\
 & DeepSeek-V4-Pro & 23.446 & 10.559 & 1230.1 & 2.220 & 1.431 \\
 & Kimi-K2.7-Code & 23.152 & 9.394 & 1228.9 & 2.465 & 1.628 \\
 & GLM-5.2 & 19.171 & 7.663 & 1291.3 & 2.502 & 1.607 \\
 & Qwen3-235B & 11.553 & 4.268 & 1159.3 & 2.707 & 1.681 \\
 & Qwen3-30B & 5.281 & 1.732 & 714.0 & 3.048 & 1.488 \\
 & Llama 4 Scout & 4.766 & 2.616 & 1576.4 & 1.822 & 1.311 \\
 & Llama 4 Maverick & 7.377 & 3.968 & 1039.0 & 1.859 & 1.308 \\
 & \emph{geometric mean} & & & & \emph{2.378} & \emph{1.510} \\
skewed & DeepSeek-V3 & 41.350 & 19.153 & 602.8 & 2.159 & 1.548 \\
 & Qwen3.5-397B & 24.885 & 9.154 & 450.4 & 2.718 & 1.860 \\
 & DeepSeek-V4-Pro & 42.728 & 20.454 & 635.0 & 2.089 & 1.469 \\
 & Kimi-K2.7-Code & 42.354 & 18.686 & 617.8 & 2.267 & 1.645 \\
 & GLM-5.2 & 36.189 & 15.886 & 622.9 & 2.278 & 1.607 \\
 & Qwen3-235B & 21.759 & 9.105 & 543.4 & 2.390 & 1.654 \\
 & Qwen3-30B & 9.708 & 3.730 & 331.6 & 2.603 & 1.412 \\
 & Llama 4 Scout & 9.539 & 5.921 & 696.3 & 1.611 & 1.253 \\
 & Llama 4 Maverick & 11.684 & 6.847 & 602.2 & 1.706 & 1.250 \\
 & \emph{geometric mean} & & & & \emph{2.173} & \emph{1.510} \\
\bottomrule
\end{tabular}
\end{footnotesize}
\end{center}
\vskip -0.1in
\end{table*}

\clearpage

\end{document}